\documentclass[english,twocolumn]{article}

\usepackage{graphicx}
\usepackage{hyperref}
\usepackage{authblk}
\usepackage{float}
\usepackage{bbold}
\usepackage{geometry}
\usepackage[backend=biber, style=numeric-comp, sorting=none]{biblatex}
\usepackage{multicol}
\usepackage{amsmath}
\usepackage{bm}
\usepackage{braket}
\usepackage[utf8]{inputenc}

\usepackage{physics}
\usepackage{color,soul}
\begin{document}

\title{High-dimensional coherent one-way quantum key distribution}

\author[1]{Kfir Sulimany}
\author[1]{Guy Pelc}
\author[2]{Rom Dudkiewicz}
\author[1]{Simcha Korenblit}
\author[1]{Hagai S. Eisenberg}

\author[1 *]{Yaron Bromberg}
\author[2 *]{Michael Ben-Or}

\affil[1]{Racah Institute of Physics, The Hebrew University of Jerusalem, Jerusalem 91904, Israel}
\affil[2]{School of Computer Science \& Engineering, The Hebrew University of Jerusalem, Jerusalem, 91904 Israel}
\affil[*]{Corresponding authors: yaron.bromberg@mail.huji.ac.il, benor@cs.huji.ac.il}

\date{}

\twocolumn[\begin{@twocolumnfalse}

\maketitle
\begin{abstract}
High-dimensional quantum key distribution (QKD) offers secure communication, with secure key rates that surpass those achievable by QKD protocols utilizing two-dimensional encoding. However, existing high-dimensional QKD protocols require additional experimental resources, such as multiport interferometers and multiple detectors, thus raising the cost of practical high-dimensional systems and limiting their use. Here, we present and analyze a novel protocol for arbitrary-dimensional QKD, that requires only the hardware of a standard two-dimensional system. We provide security proofs against individual attacks and coherent attacks, setting an upper and lower bound on the secure key rates. Then, we test the new high-dimensional protocol in a standard two-dimensional QKD system over a 40 km fiber link. The new protocol yields a two-fold enhancement of the secure key rate compared to the standard two-dimensional coherent one-way protocol, without introducing any hardware modifications to the system. This work, therefore, holds great potential to enhance the performance of already deployed time-bin QKD systems through a software update alone. Furthermore, its applications extend across different encoding schemes of QKD qudits.
\newline
\end{abstract}
\end{@twocolumnfalse}]

\section{Introduction}
Quantum key distribution (QKD) is an advanced technology that provides ultimate secure communication by exploiting quantum states of light as information carriers over communication channels \cite{bennett2000quantum,gisin2002quantum, lo2014secure, pirandola2020advances,islam2018high}. In the early QKD protocols, each bit of the key was encoded using a quantum state belonging to a two-dimensional Hilbert space \cite{bennett2020quantum,ekert1991quantum}. High-dimensional QKD protocols were introduced more recently, based on preparing a set of states belonging to a high-dimensional Hilbert space, called qudits \cite{bechmann2000quantum,cerf2002security, ali2007large, sheridan2010security,leach2012secure, bunandar2015practical}. The higher information capacity of qudits allows a higher secure key rate and improves the robustness to noise, leading to higher threshold values of the quantum bit error rate (QBER) \cite{cozzolino2019high, ecker2019overcoming}. 

Time-bin encoding of weak coherent laser pulses is the most popular technique for implementing QKD over single-mode fibers \cite{korzh2015provably,boaron2018secure, da2019experimental,bouchard2021quantum}. Recent proposals and demonstrations of high-dimensional temporal encoding showed a significant key rate improvement \cite{islam2017provably, islam2019scalable, vagniluca2020efficient,wang2021round,doda2021quantum,jachura2021photon,iqbal2020high,zhang2014unconditional,brougham2013security}. In particular, a record-breaking key rate of 26.2 Mbit/s was achieved with a four-dimensional time-bin protocol that is robust against the most general (coherent) attacks \cite{islam2017provably}. Furthermore, high-dimensional time-bin encoding was successfully demonstrated in entanglement-based QKD systems over free-space \cite{bulla2023nonlocal}, and fiber links \cite{mower2013high,zhong2015photon,lee2019large}.

Implementation of high-dimensional QKD protocols in commercial systems is still held back since present high-dimensional schemes require significantly more complex experimental resources, relative to the cost-effective two-dimensional systems \cite{bunandar2018metropolitan}. The large experimental overhead results from the fact that high-dimensional encoding not only increases the channel capacity but also increases the amount of information that Eve can extract. Most QKD protocols limit the amount of information accessible to Eve by projecting the quantum states at the receiver's end on unbiased bases. While the projection in two-dimensional schemes is usually implemented with a single interferometer followed by a single photon detector (SPD), to fully exploit the potential capacity of d-dimensional schemes, $O(d)$ imbalanced interferometers and $O(d)$ SPDs \cite{islam2017provably, islam2019scalable, vagniluca2020efficient} are required. Thus, to date, all high-dimensional QKD systems implementations require complex and expensive systems that are impractical for commercial applications. 

In this work, we present a different approach for high-dimensional QKD with time-bin encoding, which can be implemented using a standard commercial QKD system without any hardware modifications. Instead, we show that the eavesdropper’s information (Eve) can be bounded by simply randomizing the time-bins order of the qudits sent by the transmitter (Alice). Our approach is particularly relevant for systems in the detector saturation regime, where the secure key rate is limited by the number of photons that can be detected by the receiver. The detector saturation regime is the common regime among deployed systems \cite{lee2019large}.

We further analyze the expected secure key rate. We introduce two security analyses against two standard types of attacks; individual attacks, where the Eavesdropper is limited to interact with each time-bin separately and independently, and coherent attacks, which is the most powerful class of attacks. The analysis of individual attacks provides an upper bound for the secure key rate, assuming a realistic attack with currently available technologies. While analyses for two-dimensional QKD protocols have been introduced by Devetak–Winter bound \cite{devetak2005distillation, branciard2008upper}, the generalization of the analytical methods for high-dimensional protocols is elusive due to the challenge of diagonalization of the density matrices representing the states. In our analysis, we solve this issue by choosing the natural basis that divides the state's space into orthogonal subspaces. By that, we extract an analytical formula for the secure key rate upper bound. The secure key rate lower bound is obtained by generalizing the analysis in \cite{moroder2012security} which employs a numerical method to provide optimized key rates for two-dimensional systems, to higher dimensions. The complete open-source package used in this work is available online \cite{HDQKD2023}.


We experimentally demonstrate a 32-dimensional protocol over a 40 km long fiber using a standard coherent one-way (COW) QKD system that requires only two single-photon detectors and one interferometer at the receiver end. We show that our new high-dimensional protocol yields a two-fold increase in the asymptotically secure key rate for individual attacks, compared to the two-dimensional COW protocol, using the exact same experimental setup.

\section{Protocol Scheme}

Traditionally, high-dimensional quantum key distribution protocols have relied on the utilization of multiple optical modes to encode quantum states belonging to a set of two mutually unbiased bases \cite{bechmann2000quantum,cerf2002security}. By measuring the error rates associated with these bases, the extent of information that Eve might possess can be limited, as her actions must maintain the measured error rates. To measure the error rates in the two mutually unbiased bases, Bob typically employs a multi-port interferometer, often employing a discrete Fourier transform (DFT), followed by multiple single-photon detectors.

\begin{figure}[ht!]
\begin{centering}
\includegraphics[width=\columnwidth]{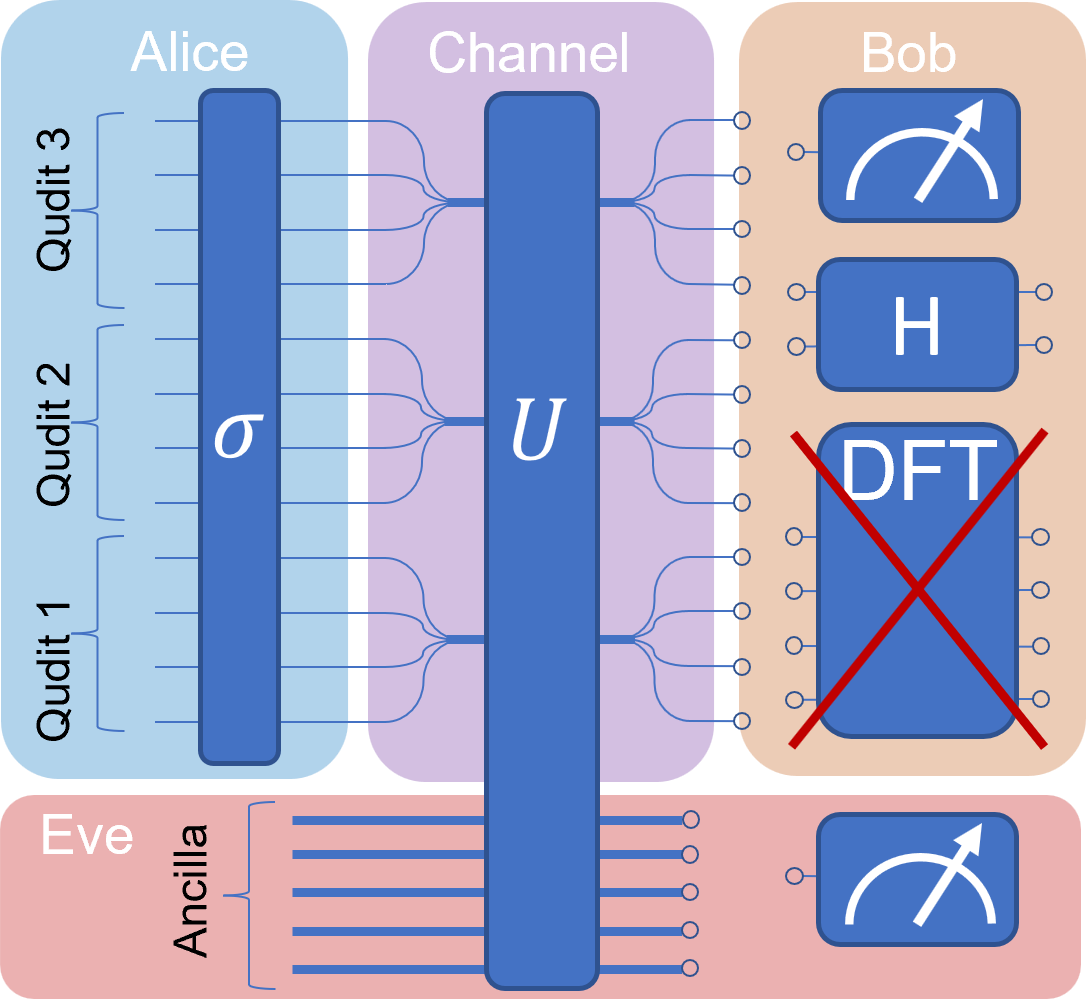}
\par\end{centering}
     \caption{ \textbf{Protocol scheme with arbitrary modes.} Alice encodes three 4-dimensional qudits, represented here using 12 dual-rail optical modes, and applies a secret random permutation ($\sigma$). Bob is restricted to measuring the photon occupation in each mode and applying a Hadamard ($H$) transformation. Importantly, since Bob does not use a discrete Fourier transform (DFT) measurement, he can use a standard two-dimensional system with a two-port interferometer. Eve attacks the quantum channel by applying a nonlocal unitary transformation $U$. The transformation $U$ operates on a Hilbert space containing the encoded qudits space and an arbitrary-dimension ancilla qudits space, while $\sigma$, $H$, and $DFT$ operate on the optical modes space.}
 \label{scheme}
 \end{figure}

In our proposed protocol, we impose limitations on the information accessible to Eve by evaluating the error rate associated with a basis defined by a set of multiple modes, together with the visibility of interference between their nearest neighbors, as depicted in Figure \ref{scheme}. The visibility measurements are conveniently obtained by employing a two-port Hadamard (H) transformation. To further restrict Eve's potential access to information, Alice randomizes the ordering of the modes using a permutation $\sigma$. Consequently, any transformation applied by Eve must preserve the relative phase between these modes. Although these constraints on Eve's actions may not be as strong as the DFT measurement, they suffice, as we show below, to limit Eve's potential knowledge.

Importantly, the key features of the described protocol do not depend on the specific set of optical modes nor on the particular photonic degrees of freedom that are used for encoding the qudits. Moreover, they are equally applicable for implementations based on weak coherent pulses, single-photon states, or entangled photon pairs. In this work, we focus on analyzing the new protocol using time-bin encoding with weak coherent pulses, in a configuration that generalizes the widely used coherent one-way (COW) protocol.

\subsection{High-dimensional coherent one-way protocol}
In COW protocol the bit string is encoded by the time of arrival of weak coherent laser pulses, and the channel disturbance is monitored by measuring the visibility of the interference between neighboring pulses \cite{stucki2005fast}. That is, bits 0 and 1 are sent using $|\alpha\rangle|0\rangle$ and $|0\rangle|\alpha\rangle$, respectively, where $|0\rangle$ is the vacuum state and $|\alpha\rangle$ is a coherent state. The receiver (Bob) simply recovers the bit value by measuring the arrival time of the laser pulse. To detect attacks, a small fraction of the pulses splits to a monitoring line by a fiber beam splitter. In the monitoring line, Bob checks for phase coherence between any two successive laser pulses by using an imbalanced interferometer and one single photon detector. \cite{gisin2004towards,stucki2009high,stucki2009continuous,walenta2014fast,korzh2015provably,sibson2017chip,roberts2017modulator, sibson2017integrated, dai2020pass}.

\begin{figure}[ht!]
\begin{centering}
\includegraphics[width=\columnwidth]{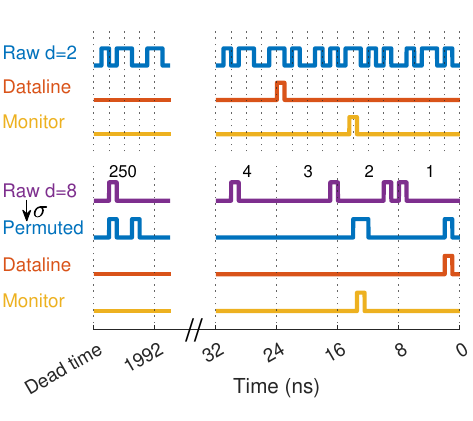}
\par\end{centering}
     \caption{ \textbf{Example of the protocol scheme implemented with time-bin encoding.} In 2-dimensional encoding ($d=8$), Alice transmits $n=1000$ two-dimensional time-bin encoded signals within a time window of 2000 ns, which corresponds to the detector's dead time. At the data line, Bob measures one photon per dead time, yielding 1 bit of mutual information. Additionally, at the monitoring line, Bob measures one photon per dead time, corresponding to a coherence test of one pair of pulses. In 8-dimensional encoding  ($d=8$), Alice transmits $n=250$ signals per dead time to Bob, after permuting the time-bins order. Alice may apply the permutation in the software before the generation of the optical pulses. In the data line, Bob measures one photon per dead time yielding 3 bits of mutual information after Alice reveals the permutation. Irrespective of the encoding dimension, Bob employs the same monitoring line to detect potential attacks. }
 \label{Scheme_time-bins}
 \end{figure}

In our high-dimensional protocol, Alice encodes the qudits of the raw key by sequences of $d$ time bins. In each sequence, only a single time bin is populated. She groups $n$ sequences to a block and applies a random permutation on the entire block, creating a series of randomly occupied time bins (the \textit{permuted key} in Fig. \ref{Scheme_time-bins}). Next, Alice transmits to Bob a series of pulses that occupy the time-bins according to the \textit{permuted key}). The permutation can then be announced by a public channel so that Bob can recover the raw key by applying the inverse permutation on the time bins measured at the data line. 

In two-dimensional COW protocols, the security stems from testing the coherence between successive pulses in Bob's monitoring line. In principle, in high-dimensional COW protocols, testing the phase coherence only between successive pulses is insufficient, because standard security proofs for high-dimensional QKD protocols require testing the phase coherence of multiple time-bins. Nevertheless, thanks to the permutation that Alice applies, any two pulses in the raw key have a finite probability of being mapped to two successive pulses in the permuted key. Eve, therefore, must preserve the relative phase between any two successive pulses. As we show below, this is sufficient for quantifying the amount of information Eve can extract and setting bounds on the secure key rate.

Formally, let $ q_{0},...,q_{n-1} \in\{1 ... d\}$ be the raw key Alice wants to transmit. Alice chooses a random permutation $\sigma$ of $\{1...d\cdot n\}$ and over the next $d\cdot n$ time bins sends a coherent state $|\alpha\rangle$ at time bin $t$ if $t=\sigma(d\cdot i + q_i)$, where $i$ in $\{0...n-1\}$ is the signal index, and a vacuum state $|0\rangle$ otherwise (see Fig. \ref{Scheme_time-bins}). After Bob measures the pulse sequence, Alice transmits the permutation $\sigma$ over the classical channel. When Bob detects a click at time $t$, he calculates $\sigma ^{-1}(t) = i\cdot d + j$ for $i \in \{0...n-1\}$ and $j \in \{1...d\}$. Bob then announces over the public channel the signal index $i$ at which he detected a photon to generate a sifted key with Alice. 
The number of bits in the sifted key, per detector dead time, is thus $log_2(d)$. High-dimensional encoding, therefore, has a clear advantage in the detector-saturated regime. Nevertheless, the amount of information that is revealed to an eavesdropper may also increase with the dimension $d$. The security proof of the protocol thus relies on finding the optimal dimension that maximizes the final secure key rate.  

\section{Security Analysis} \label{Security Analysis}

Our high-dimensional security analysis is based on the analytical framework developed for the standard two-dimensional COW protocol. An upper bound on the secure key rate of two-dimensional COW protocols was derived by analyzing individual attacks \cite{branciard2008upper}. It was found that the upper bound scales linearly with the transmittance of the channel. Later, lower bounds on the secure key rate against general attacks have been derived, showing quadratic scaling \cite{moroder2012security}. However, so far, all COW-QKD systems \cite{ gisin2004towards,stucki2009high,stucki2009continuous,walenta2014fast,korzh2015provably,sibson2017chip,roberts2017modulator, sibson2017integrated, dai2020pass}, still employ the original security proof \cite{branciard2008upper} since it is the most realistic attack with current technologies. More recently, zero-error attacks, an eavesdrop without breaking the coherence between
adjacent non-vacuum pulses, have also been studied for the COW protocol \cite{trenyi2021zero,curty2021foiling,branciard2006zero}. These attacks require an extremely low number of detected photons and are thus irrelevant to the detector-saturated regime that we consider in this work.

\subsection{Secure key rate upper bound} \label{chapter upper bound}

To calculate the upper bound of the secure key rate for individual attacks, we focus on Eve's action on a single time bin. It can be defined by a linear transformation describing the action on non-occupied (vacuum) and occupied (coherent state) time bins $|0\rangle_{A},|\sqrt{\mu}\rangle_{A}$:

\begin{equation} \label{eq1}
|0\rangle_{A}|\varepsilon\rangle_{E} \to |0\rangle _{B} |v_{0}\rangle _{E} + \sqrt{Q \mu t}|1\rangle _{B} |p_{0}\rangle _{E}
\end{equation}

\begin{equation} \label{eq2}
|\sqrt{\mu}\rangle_{A} |\varepsilon\rangle_{E} \to |0\rangle _{B} |v_{\mu}\rangle _{E} + \sqrt{(1-(d-1)Q)\mu t}|1\rangle _{B} |p_{\mu}\rangle _{E},
\end{equation}
where $|v_j\rangle_E$ $(j \in \{0, \mu\})$ are the states that Eve attaches to the vacuum part of the signal,
while $|p_{\mu}\rangle _{E}$ is the state that Eve attaches to the photon part of the signal. We assume $\mu t\ll1$ so that we can neglect multiphoton terms. Here we assume Eve's states can be arbitrary, conditioned to some constraints described below. The probability amplitude of each of the terms guarantees that Eve's attack does not increase the quantum bit error rate $Q$ (QBER) of the channel, i.e. the probability that Bob receives a wrong bit value.

Coherence between occupied time bins is monitored by analyzing the detection events in the monitoring line. The phase delay between the two arms in the monitoring line is chosen such that two successive non-empty pulses sent by Alice will interfere destructively in one output port and constructively in the other port. We quantify the degree of coherence by the visibility:

\begin{equation} \label{eq3}
V = \frac{P(D_c)-P(D_d)}{P(D_c)+P(D_d)},
\end{equation}
where $P(D_c)$ and $P(D_d)$ are the probabilities to measure a photon at the constructive and destructive ports, respectively. Since Eve's attack should conserve visibility, we can derive a constraint on Eve's action on two successive pulses sent by Alice $|\sqrt{\mu}\rangle|\sqrt{\mu}\rangle$. Assuming $\mu t \ll 1$,  Eve's action is given by:

\begin{equation} \label{eq4}
\begin{gathered} 
|\sqrt{\mu},\sqrt{\mu}\rangle_A |\varepsilon,\varepsilon\rangle_{E} \to
|0,0\rangle_B |v_\mu ,v_\mu \rangle _E \\
+ \sqrt{(1-(d-1)Q)\mu t}[|1,0\rangle_B |p_\mu ,v_\mu \rangle _E   \\
+|0,1\rangle_B |v_\mu ,p_\mu \rangle _E],
\end{gathered}
\end{equation}

The visibility constraint then yields (see Supplementary Information \ref{Sup_Detailed security analysis}):

\begin{equation} \label{eq5}
\begin{gathered} 
V=|_E\langle p_{\mu}|v_{\mu}\rangle_E|^{2}. 
\end{gathered}
\end{equation}

The last constraint on Eve's transformation is that it must be unitary. Thus in the $\mu t \ll 1$ limit we get from Eq. \ref{eq1} and \ref{eq2}:

\begin{equation} \label{eq6}
\begin{gathered} 
_E\langle v_0 | v_\mu \rangle_E = e^{- \mu /2} 
\end{gathered}
\end{equation}

Our security analysis is therefore based on three constraints imposed on Eve's action: i) It must retain the QBER (eq. \ref{eq1}), ii) it must conserve the visibility (eq. \ref{eq5}), and iii) it must be unitary (eq. \ref{eq6}).

To compute the amount of information that can be extracted by Eve, as quantified by the Holevo information \cite{cabello2000quantum}, we first need to analyze her action on a qudit with occupation $\mu^{(i)}$ in the $i^{th}$ time bin. Neglecting all multi-photon terms, Eve's action can be presented by:
\begin{equation} \label{eq7}
\begin{gathered}
|0,...,0,\sqrt{\mu}^{(i)},0,...,0\rangle_A|\varepsilon,...,\varepsilon\rangle_E \to \\
|0,...,0\rangle_B \otimes |V_{i}\rangle \\
+  \sqrt{(1-(d-1)Q)\mu t}|0,...,0,1^{(i)},0...,0\rangle_B \otimes |C_{i}\rangle\\
+ \sum_{k=1...d, k\neq i}\sqrt{Q \mu t} |0,...,0,1^{(k)},0...,0\rangle_B |W_{i,k}\rangle,
\end{gathered}
\end{equation}
where $|V_{i}\rangle = |v_{0},...,v_{0},v_{\mu}^{(i)},v_{0},...,v_{0}\rangle_E$ is Eve's state representing the case where she sends a vacuum state at time bin $i$, $|C_{i}\rangle = |v_{0},...,v_{0},p_{\mu}^{(i)},v_{0},...,v_{0}\rangle_E$ is Eve's state representing the case where she sends to Bob a photon at the right time bin $i$, and $|W_{i,k}\rangle = |v_{0},...,v_{0},p_{0}^{(k)},v_{0},...,v_{0},v_{\mu}^{(i)},v_{0},...,v_{0}\rangle_E$ is Eve's state representing the case where she sends to Bob a photon at the wrong time bin $k$.

Next, we compute the density matrices of Eve's subsystem, conditioned by the event where Alice sends a pulse at time bin $i$ and Bob detects a photon at some arbitrary time bin: 
\begin{equation} \label{eq8}
\begin{gathered}
\rho_{E}^{A=i} = (1-(d-1)Q)|C_{i}\rangle \langle C_{i}| \\ 
+ \sum_{k=1...d, k\neq i}Q |W_{i,k}\rangle \langle W_{i,k}| 
\end{gathered}
\end{equation}
Similarly, the density matrix of Eve's subsystem is conditioned by the event where Bob detects a pulse at time bin $i$ and Alice sent the pulse at an arbitrary time bin:
\begin{equation} \label{eq9}
\begin{gathered}
\rho_{E}^{B=i} = (1-(d-1)Q)|C_{i}\rangle \langle C_{i}| \\
+ \sum_{k=1...d, k\neq i}Q |W_{k,i}\rangle \langle W_{k,i}|
\end{gathered}
\end{equation}

 Knowing these density matrices, we can apply the Devetak–Winter bound for the secure key rate by calculating the Holevo information \cite{devetak2005distillation}. The Holevo information on Alice-Eve channel $\chi_{AE}$ and on Bob-Eve channel $\chi_{BE}$, are defined by:
\begin{equation} \label{eq10}
\begin{gathered}
\chi_{AE}  = S\left(\sum_{i=1}^{d} \frac{1}{d}\rho_{E}^{A=i}\right) - \sum_{i=1}^{d}\frac{1}{d}S\left( \rho_{E}^{A=i}\right) \\
\chi_{BE}  = S\left(\sum_{i=1}^{d} \frac{1}{d}\rho_{E}^{B=i}\right) - \sum_{i=1}^{d}\frac{1}{d}S\left( \rho_{E}^{B=i}\right)
\end{gathered}
\end{equation}
where $S(\rho)=-Tr(\rho\log_{2} \rho)$ is the von Neumann entropy of $\rho$. The maximal information Eve can extract is bounded by the maximum of these two quantities.  Direct computation of eq. \ref{eq10} shows that $\chi_{BE}>\chi_{AE}$ (see Supplementary Information \ref{Sup_Detailed security analysis}), hence from this point on, we will focus on analyzing $\chi_{BE}$.

 Eve has no constraints over $|p_{0}\rangle$ as it does not affect the three constraints imposed by eqs.(\ref{eq1}),(\ref{eq2}),(\ref{eq5}) and (\ref{eq6}). Thus, in order to maximize her information she can choose $|p_{0}\rangle$ that is orthogonal to all other vectors $|v_{0}\rangle,|v_{\mu}\rangle,|p_{\mu}\rangle$. Conveniently, we can then separate the trace of the above matrices to  a trace of density matrices that contain only one $|p_{0}\rangle$ in time bin $i$ for $i=1..d$, and a trace of matrices that do not contain $|p_{0}\rangle$,  yielding (see Supplementary Information \ref{Sup_Detailed security analysis}):
 
 \begin{equation} \label{eq11}
\begin{gathered}
\chi_{BE}= \\
\sum_{k=1}^{d} \left[S\left(\sum_{i=1}^{d-1} \frac{Q}{d}  |W'_{i}\rangle \langle W'_{i}| \right)\right. 
\left. -\frac{1}{d} S \left(\sum_{i=1}^{d-1}Q|W'_{i}\rangle \langle W'_{i}| \right)
\right] \\ 
+ S\left(\sum_{i=1}^{d-1} \frac{1-(d-1)Q}{d} |C_{i}\rangle \langle C_{i}| \right) \\
-\frac{1}{d}\sum_{i=1}^{d} S \left(\left(1-(d-1)Q\right)|C_{i}\rangle \langle C_{i}| \right) 
\end{gathered}
\end{equation}

where we define $|W'_{i}\rangle = \overbrace{|v_{0},...,v_{0},v_{\mu}^{(i)},v_{0},...,v_{0}\rangle}^{d-1\ terms}$, such that $|W'_{i}\rangle \otimes |p_0\rangle = |W_{i,d}\rangle$ and $|W_{i,j}\rangle$ are equivalent up to reordering the order of the time bins.

After we diagonalized the density matrices and computed  traces, we obtained the following expression for the Holevo information (See Supplementary Information \ref{Sup_Detailed security analysis}):

 \begin{equation} \label{eq12}
\begin{gathered}
\chi_{BE}  = Q(d-1)\log_{2}(d) \\
+ S\left( \frac{1-(d-1)Q}{d} \left((d-1)|\langle v_0 | p_\mu \rangle| ^{2} +1 \right) \right) \\
+ (d-1) S\left( \frac{1-(d-1)Q}{d} \left(1-|\langle v_0 | p_\mu \rangle| ^{2} \right) \right) \\
- S \left(1-(d-1)Q \right)
\end{gathered}
\end{equation}

To maximize eq. $\chi_{BE}$, we can minimize $\langle v_0 | p_\mu \rangle$ under the constraints imposed by eq. \ref{eq5} and eq. \ref{eq6}. Using a parametric representation of $|v_0\rangle,|p_u\rangle$ and $|v_u\rangle$ in 3-D space, we find that the maximal information Eve can extract $\max\{\chi_{BE}\}$ is obtained for:

\begin{equation} \label{eq13} 
\begin{gathered}
\langle v_0 | p_\mu \rangle = e^{-\mu/2}\sqrt{V}-\sqrt{ 1-e^{-\mu} } \sqrt{1-V}.
\end{gathered} 
\end{equation}
 
An upper bound on the secure key fraction can now be computed, using the bound \cite{branciard2008upper}.
\begin{equation} \label{upper_bound}
\begin{gathered}
I_{AB} = \log_{2}(d) + (d-1)Q\log_{2}(Q)+ \\ \left(1-(d-1)Q\right)\log_{2}\left(1-(d-1)Q\right) \\
- \max\{\chi_{BE}\}.
\end{gathered}
\end{equation}

We compare the secure key rates and analyze the resilience to noise in the supplementary material \ref{Reselience to noise}. We find that while increasing the dimensional encoding might increase the secure key rate for low error rates, the resilience to noise does not improve with dimensional encoding, see Figure \ref{Res_to_noise}.

\subsection{Secure key rate lower bound}

To analyze the security of the protocol against general collective attacks in the asymptotic limit, we extend the method proposed by Moroder et al \cite{moroder2012security} for high-dimensional encoding, which employs a numerical method to provide optimized key rates \cite{winick2018reliable, hu2022robust, araujo2023quantum}. 

For completeness, in this chapter we present the security analysis derived in \cite{moroder2012security}, with the necessary adaptation for high dimensional encoding. The adaptation includes the use of the secure rate formula generalized for high dimensions, as well as the generalized high-dimensional positive operator value measurements (POVM) and the sifting maps acting on high-dimensional states. In addition, the conversion from infinite dimensional space representing coherent states to a finite dimension that can be numerically simulated is also been generalized to high dimensional encoding in this work. Lastly, we have taken into account the dead time of the detectors, resulting in the detectors entering a saturation regime. The complete open-source Matlab package used in this work is available online \cite{HDQKD2023}.

General attacks often do not have an advantage over collective attacks when the de Finetti theorem applies \cite{renner2008security}. The theorem holds when the quantum state shared by Alice and Bob is composed of blocks and the state itself is invariant under permutations of the blocks' order. 

The problem with the coherent one-way protocol is that it induces a fixed ordering of the signals by coherent measurements between successive time bins. Therefore, the state is not invariant under the permutation of the signals. This issue can be solved by grouping time bins into blocks of size $d$, separated by an empty time bin. When permuting these blocks, the coherence within the blocks is preserved. Therefore, the security against collective attacks for a block implies security against coherent attacks in the original settings. We emphasize that although adding an empty time bin decreases the transmission rate of information in the channel, in the regime of a large dead time this effect is negligible.

\subsubsection{Secure key rate formula}

Let $\rho_{ABE}$ represent the density matrix of the state shared by all parties after transmitting a block signal from Alice to Bob. Alice and Bob generate a sifted key based on a public announcement denoted by $v$. This generation process can be presented using maps $\Lambda_{v}^A\otimes\Lambda_{v}^B$. The three parties share the state $\sigma_{\Bar{A}\Bar{B}E}$ which is determined by $\Lambda_{v}^A\otimes\Lambda_{v}^B(\rho_{ABE}) = p(v)\sigma_{\Bar{A}\Bar{B}E}$, where announcement $v$ occurs with a probability of $p(v)$, and $\Bar{A}$ and $\Bar{B}$ represent the output spaces, which serve as logical qudits.

The one-way classical post-processing key rate formula, described in \cite{tomamichel2011uncertainty}, can be used to extract a lower bound on the secure key rate for every announcement $v$. This lower bound is given by $\log_2(d)-h_d(e_v)-h_d(\delta_v)$, where $h_d(x) = -x\log\frac{x}{d-1} - (1-x)\log(1-x)$ represents the d-dimensional entropy. Here, $e_v$ represents the bit error rate between, and $\delta_v$ represents the phase error that would be obtained if both Bob and Alice measured in a mutually unbiased basis. The estimation of $\delta_v$ is essential for establishing an upper bound on Eve's knowledge of the sifted key. In practice, it is often more convenient to consider the averaged value of conclusive announcements $v\in V_c$ and the total probability of a conclusive measurement $G=\sum_{v\in V_c}p(v)\leq 1$. The secure key rate lower bound per block is obtained:

\begin{equation} \label{coherent_1}
\begin{gathered} 
R_m \geq \inf_{\rho_{ABE}}\sum_{v\in V_c}p(v)[\log_2(d)-h_d(e_v)-h_d(\delta_v)] \\
\geq \inf_{\rho_{ABE}}G[\log_2(d)-h_d(\Bar{e}_c)-h_d(\Bar{\delta})] \\
\geq G[\log_2(d)-h_d(\Bar{e}_c)-h_d(\Bar{\delta}^{max})]
\end{gathered}
\end{equation}

In order to bound the value of $R_m$ from below, we utilized the concavity property of $h_d$ and expressed it in terms of the averaged error rates $\Bar{e}_c=\sum_{v\in V_c}p(v)e_v/G$ and $\Bar{\delta}=\sum_{v\in V_c}p(v)\delta_v/G$. It is important to note that $\Bar{e}_c$ and $G$ are observable quantities, whereas $\Bar{\delta}$ is not directly measured in our protocol. Therefore, we estimate $\Bar{\delta}^{max}$, which represents the largest phase error that satisfies all other constraints. To maximize the function $h_d$ within the range $[0,1-1/d]$, we can search for the maximum value of $\Bar{\delta}^{max}$, as $h_d$ is a monotonically increasing function in this range.

\subsubsection{Phase error estimation}
A significant difficulty in calculating Eq. \ref{coherent_1} lies in establishing an upper limit for the average phase error, denoted as $\Bar{\delta}$. This parameter can be expressed as the expectation value on the original bipartite state  $\rho_{AB}=tr_E(\rho_{ABE})$, utilizing the maps $\Lambda_{v}^A\otimes\Lambda_{v}^B$:

\begin{equation} \label{coherent_2}
\begin{gathered} 
\Bar{\delta} = \sum_{v\in V_c}p(v)tr(\sigma_{\Bar{A}\Bar{B},v}F_{\delta_v})
=\sum_{v\in V_c}tr[\Lambda_{A,v}\otimes\Lambda_{B,v}(\rho_{AB})F_{\delta_v}] \\
= tr[\rho_{AB}\sum_{v\in V_c}\Lambda^{A\dag}_{v}
\otimes\Lambda^{B\dag}_{v}(F_{\delta_v})]
= tr(\rho_{AB}F_{\Bar{\delta}})
\end{gathered}
\end{equation}
The operators $F_{\delta_v}$ represent the corresponding phase error operators, see supplemetary informatiom \ref{SUP sifting}

When Alice and Bob possess partial knowledge about the state $\rho_{AB}$, it can be represented as known expectation values $k_i= tr(\rho_{AB}K_i)$ for specific operators $K_i$, which are constructed using local projection measurements at the sites of Alice and Bob. In the supplementary information, we define the set $K_i$, (see \ref{Sup_Measurement_Description}). To calculate the outcome probabilities $k_i$, we evaluate $tr(\Tilde{\rho}_{AB}K_i)=k_i$, where $\Tilde{\rho}_{AB}$ represents the bipartite state of Alice and Bob without an eavesdropper, but accounting for the losses and errors resulting from channel imperfections (see \ref{Channel model}).

To find the maximum phase error $\Bar{\delta}^{max}$ under the constraints, we define the semidefinite program:

\begin{equation} \label{coherent_3}
\begin{gathered} 
\bar{\delta}^{max} = \max tr(\rho_{AB}F_{\Bar{\delta}}) \\
s.t. \ \rho_{AB}\geq0, \  tr(\rho_{AB}K_i)=k_i \  \forall i.
\end{gathered}
\end{equation}
The convex optimization problem can be efficiently solved using standard numerical tools, enabling us to obtain the precise optimal value of $\Bar{\delta}^{max}$. It is worth noting that optimizing the linear objective $\Bar{\delta}$ rather than $\sum_{v\in V_c}p(v)[h_d(\delta_v)]$ enables the formulation of a linear optimization problem. However, this approach does not provide a tighter lower bound. Specifically, the lower bound obtained through the linear problem will be less tight for higher dimensions.

\subsubsection{Numerical optimization}

\begin{figure}[ht!]
\begin{centering}
\includegraphics[width=\columnwidth]{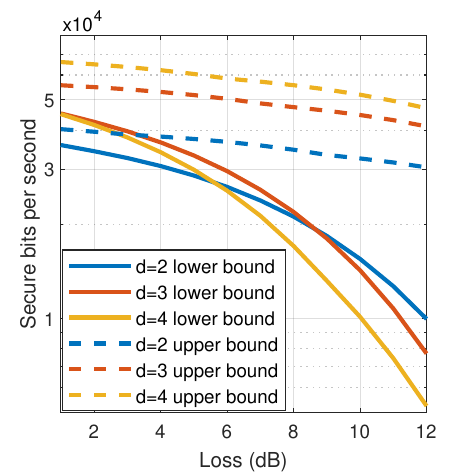}
\par\end{centering}
     \caption{ \textbf{Bound of the secure key rate for different dimensions.} The lower bound (solid lines) and the upper bound (dashed lines) on the secure key rate on a logarithmic scale are presented versus the total system loss in dB. Different colors represent different dimensional encoding sizes. A finite secure key rate lower bound is achieved for dimension $2,3,4$, indicating that the protocol is secure for these dimensions. The optimal lower bound secure key rate is achieved for $d=3$ in the range of up to 8 dB loss. The upper bound is of the same order of magnitude as the lower bound, indicating that it represents an effective attack. Here Bob’s detector dark count rate is $5\cdot10^{-8}$, bit error rate of $0.5\%$, visibility of $99.5\%$, dead time of $10us$, and a pulse duration of $2ns$.}
 \label{Fig secure bits per second}
 \end{figure}

First, we study the lower bound of the secure bits per block, obtained by equation \ref{coherent_1}. For a given total system loss, that includes channel losses and Bob’s finite detection efficiency, we search for the optimal signal occupation $\mu$ that maximizes the lower bound. As expected, we find that our high-dimensional encoding decreases the secure key per block, which exhibits a quadratic decrease with the loss for any dimensional encoding size, see supplementary information \ref{Lower bound of the secure key per block}. The secure bits per second, however, increase with the dimensionality up to an optimal dimension (see supplementary information section \ref{sec:rate_per_second_analysis}). For example, with parameters corresponding to a typical system: dark count rate of $5\cdot10^{-8}$, bit error rate of $0.5\%$, visibility of $99.5\%$, dead time of $10us$, and a pulse duration of $2ns$, the optimal secure key per second is achieved for d = 3 in the range of loss up to 8.5dB (Figure \ref{Fig secure bits per second}). 

To estimate the tightness of the lower bound found by the numerical optimization, we compare it to the upper bound, which is computed by multiplying the number of secure bits per photon calculated using Eq. \ref{upper_bound} by the number of photons per second. The obtained curves depicted in figure \ref{Fig secure bits per second} show that both bounds are of the same order of magnitude.

\section{Experimental Implementation}
\begin{figure}[ht!]
\begin{centering}
\includegraphics[width=\columnwidth]{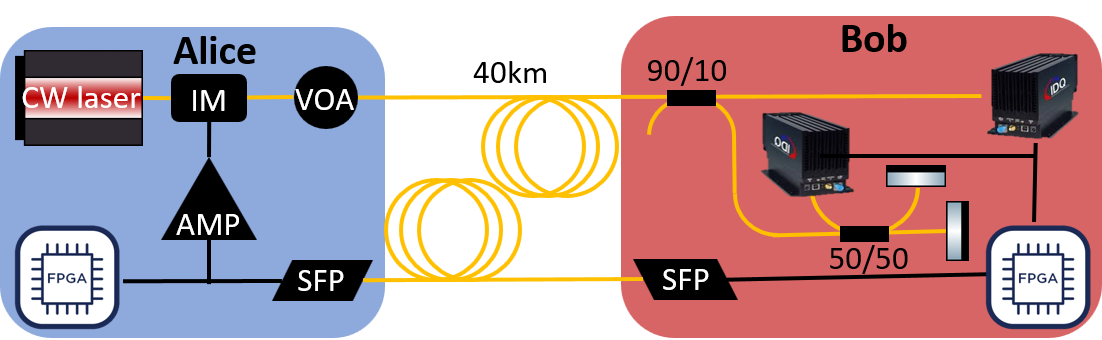}
\par\end{centering}
     \caption{ \textbf{Experimental setup for comparing arbitrary-dimensional QKD schemes.} Alice’s transmitter (left) consists of a continuous wave (CW) laser at $\lambda=1550 nm$ that is modulated using an electro-optic intensity modulator (IM) running at $500 MHz$. The pulses are passed through a variable optical attenuator (VOA) that regulates the mean photon number per pulse. The weak coherent pulses are delivered to Bob's end through a $40 km$ long single-mode fiber (SMF-28). Bob’s receiver (right) consists of an asymmetric beamsplitter, which provides a passive choice of the measurement basis; $90\%$ of the photons travel directly to the data detector, and $10\%$ pass through an unbalanced interferometer and are detected by the monitor detector. We lock the laser's wavelength to the interferometer so that the monitor detector always measures the dark port of the interferometer. The interference visibility is estimated by registering the detection events due to the interfering and non-interfering events. In addition to the $40 km$ long quantum channel that delivers the weak coherent pulses, we use a separate $40 km$ SMF-28 fiber for all classical communication between Alice and Bob and to distribute an optical clock signal between them based on the White Rabbit protocol \cite{lipinski2011white}. State preparation and sifting are run by two field-programmable gate arrays (FPGA) at Alice's and Bob's ends.}
 \label{Fig setup}
 \end{figure}

The important feature of our high-dimensional protocol is that it can be implemented in a standard two-dimensional COW system, illustrated in Fig. \ref{Fig setup}, without any hardware changes. The system consists of a transmitter (Alice) and a receiver (Bob). The transmitter sends a train of weak coherent pulses that are prepared from a continuous wave (CW) laser emitting at $\lambda=1550 nm$, by an intensity modulator (IM) running at $500 MHz$. Before leaving the transmitter, the pulses are attenuated to reach a single photon level using a  variable optical attenuator (VOA). To generate $200 ps$ long pulses with random occupations of $\tau = 2 ns$ long time-bins we use a field programmable gate array (FPGA). Synchronization is achieved over the 40 km fiber channel using the White Rabbit protocol \cite{lipinski2011white}. To interfere with successive pulses at the receiver's end, an unbalanced fiber-interferometer is installed, where we use Faraday mirrors to compensate for random polarization drifts in the fiber (Figure \ref{Fig setup}). We use single-photon avalanche detectors (SPADs) with $20\%$ detection efficiency and $400 ps$ timing resolution. The detectors’ dead time is $~4 \mu s$, limiting the maximal raw key rate to $250 kHz$.

We analyze the protocol's performance for different dimensions and compare the experimental results obtained with our QKD system with the predictions of a cleaner theoretical model. The model assumes the error rate per time bin and the visibility are independent of the dimensionality. We present the secure bits per second, obtained by multiplying the raw bits per photon by the number of detected photons per second (see supplementary information \ref{Key rates for different dimensions}). It is evident that an optimal secure bit rate is achieved for $d=8$, resulting in more than a two-fold increase in the secure bit rate for both the experimental data and the theoretical model.

While the experimental results and theoretical model exhibit similar trends, the model fails to capture the exact secure bit rate, due to the oversimplification of the model that assumes linear scaling of the QBER with the dimension size. For a description of the noise sources and the Qber and visibility measurement as a function of the dimension, see supplementary information \ref{Sup_Experimental_details}. 

\begin{figure}[ht!]
\begin{centering}
\includegraphics[width=\columnwidth]{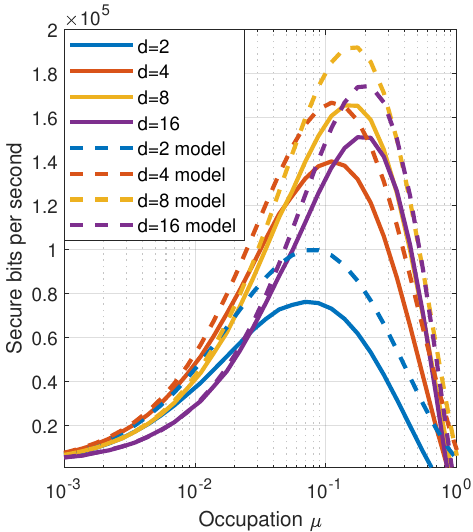}
\par\end{centering}
     \caption{ \textbf{Secure key rates for different dimensions.} Secure bits per second is achieved by multiplying the raw bits per photon by measured photons per second. The secure bits per second is obtained by measuring the bit error rate, visibility, and loss, and applying the secure key rate upper bound found in Eq. \ref{upper_bound}. Solid lines represent the result obtained in our experimental system and dashed lines represent our model system. The model assumes the error rate per time bin and the visibility are independent of the dimensionality. The optimum is achieved at $d=8$, where the number of secure bits per second increases by $2.04$ for our system, and by $1.89$ for the model.}
 \label{exp}
 \end{figure}

\section{Discussion}

By employing a novel scheme for arbitrary-dimensional QKD that leverages standard commercial hardware, we managed to enhance the secure key rate, showcasing the effectiveness of our method. The potential applicability and implications of our research, however, go well beyond our system.

First, our technique has the potential to be integrated with other time-bin QKD systems. While we showed enhancement in the secure key rate for the most popular and cost-effective commercial design based on standard single photon avalanche photodiode, a similar analysis shows an improvement also for high-end QKD systems based on superconducting nanowire detectors. For example, considering detector deadtime of $10 ns$ and modulation rates as high as $10 Gbps$, the secure bit rate at dimension $d=8$ may increase by a factor of 2 compared with standard two-dimensional COW encoding. In general, for time-bins systems, our approach for encoding high-dimensional information enables higher key rates in the detector's saturation regime. This regime is obtained in typical systems up to 100km fiber links \cite{lee2019large}. 

Moreover, our high-dimensional QKD method is not limited to time-bin encoding and is directly applicable to other QKD qudit encoding schemes, such as in the spatial and spectral degrees of freedom. For example, QKD with states encoded with spatial modes such as non-overlapping Gaussian beams, orbital angular momentum, or waveguide modes requires a multiport interferometer. Our approach may offer new opportunities for implementing high-dimensional QKD with spatial encoding relying on a single Michelson interferometer \cite{mirhosseini2015high,canas2017high,sit2017high, bouchard2018experimental,bouchard2019quantum, sulimany2022all, lib2022processing}. Similarly, applying our protocol in the spectral domain using frequency bins and a single acoustic optic modulator to interfere only with adjacent spectral modes can pave the way for high-dimensional QKD with frequency-bins \cite{reimer2016generation, kues2019quantum}. This suggests that our method could serve as a more generalized solution for high-dimensional QKD, applicable across diverse platforms and systems.

Lastly, it is worth noting the potential of our approach to apply to systems working with entangled photon pairs, in addition to weak coherent states. The adaptability of our method could further secure communication in systems exploiting entangled states with different encoding degrees of freedom including time-energy \cite{thew2004bell,brendel1999pulsed,jha2008exploring,maclean2018direct}, angle-angular momentum \cite{vaziri2002experimental,leach2010quantum,sit2017high,erhard2018twisted}, and position-momentum \cite{howell2004realization,schaeff2015experimental,wang2018multidimensional,srivastav2022quick}.

\section{Conclusion}

In conclusion, we presented a new arbitrary high-dimensional QKD protocol, supported by a security proof, which has the advantage of requiring only standard two-dimensional QKD hardware. 

We present security proofs against two classes of attacks; individual attacks and coherent attacks. We experimentally tested the protocol with a standard two-dimensional encoding system, and its performance was tested for several dimension sizes. We demonstrated more than a two-fold enhancement of the secure key rate in the saturation regime of APD detectors. Our demonstration proves that high-dimensional quantum systems allow a significant improvement in the key generation rate as compared with the two-dimensional-encoding case. At the same time, no additional hardware is required to fully implement the protocol in standard two-dimensional systems. 

Our research on high-dimensional coherent one-way quantum key distribution by interfering just neighboring modes represents a significant step toward the practical implementation of high-dimensional QKD. Our scheme, leveraging commercially available hardware, not only enhances secure key rates but also provides a potential platform for a broader application across various QKD schemes and systems. Our method's potential applications extend across other time-bin designs, different encoding of QKD qudits, and entangled photon pairs systems. Therefore, our method can act as a robust, versatile, and efficient enhancement to QKD.

\medskip
\textbf{Acknowledgements} 
This research was supported by the \textit{United States-Israel Binational Science Foundation (BSF)} (Grant No. 2017694), \textit{ISF-NRF Singapore joint research program} (Grant No. 3538/20). and by the \textit{Israel Science Foundation (ISF)} grant No. 2137/19. KS and YB acknowledge the support of the Israeli Council for Higher Education, the Israel National Quantum Initiative (INQI), and the Zuckerman STEM Leadership Program. 

\printbibliography

@article{bennett2000quantum,
  title={Quantum information and computation},
  author={Bennett, Charles H and DiVincenzo, David P},
  journal={nature},
  volume={404},
  number={6775},
  pages={247--255},
  year={2000},
  publisher={Nature Publishing Group}
}

@article{gisin2002quantum,
  title={Quantum cryptography},
  author={Gisin, Nicolas and Ribordy, Gr{\'e}goire and Tittel, Wolfgang and Zbinden, Hugo},
  journal={Reviews of modern physics},
  volume={74},
  number={1},
  pages={145},
  year={2002},
  publisher={APS}
}

@article{pirandola2020advances,
  title={Advances in quantum cryptography},
  author={Pirandola, Stefano and Andersen, Ulrik L and Banchi, Leonardo and Berta, Mario and Bunandar, Darius and Colbeck, Roger and Englund, Dirk and Gehring, Tobias and Lupo, Cosmo and Ottaviani, Carlo and others},
  journal={Advances in Optics and Photonics},
  volume={12},
  number={4},
  pages={1012--1236},
  year={2020},
  publisher={Optical Society of America}
}

@book{islam2018high,
  title={High-rate, high-dimensional quantum key distribution systems},
  author={Islam, Nurul T},
  year={2018},
  publisher={Springer}
}

@article{bennett2020quantum,
  title={Quantum cryptography: Public key distribution and coin tossing},
  author={Bennett, Charles H and Brassard, Gilles},
  journal={arXiv preprint arXiv:2003.06557},
  year={2020}
}

@article{ekert1991quantum,
  title={Quantum cryptography based on Bell’s theorem},
  author={Ekert, Artur K},
  journal={Physical review letters},
  volume={67},
  number={6},
  pages={661},
  year={1991},
  publisher={APS}
}

@article{bechmann2000quantum,
  title={Quantum cryptography using larger alphabets},
  author={Bechmann-Pasquinucci, Helle and Tittel, Wolfgang},
  journal={Physical Review A},
  volume={61},
  number={6},
  pages={062308},
  year={2000},
  publisher={APS}
}

@article{cerf2002security,
  title={Security of quantum key distribution using d-level systems},
  author={Cerf, Nicolas J and Bourennane, Mohamed and Karlsson, Anders and Gisin, Nicolas},
  journal={Physical review letters},
  volume={88},
  number={12},
  pages={127902},
  year={2002},
  publisher={APS}
}

@article{sheridan2010security,
  title={Security proof for quantum key distribution using qudit systems},
  author={Sheridan, Lana and Scarani, Valerio},
  journal={Physical Review A},
  volume={82},
  number={3},
  pages={030301},
  year={2010},
  publisher={APS}
}

@article{leach2012secure,
  title={Secure information capacity of photons entangled in many dimensions},
  author={Leach, Jonathan and Bolduc, Eliot and Gauthier, Daniel J and Boyd, Robert W},
  journal={Physical Review A},
  volume={85},
  number={6},
  pages={060304},
  year={2012},
  publisher={APS}
}

@article{cozzolino2019high,
  title={High-Dimensional Quantum Communication: Benefits, Progress, and Future Challenges},
  author={Cozzolino, Daniele and Da Lio, Beatrice and Bacco, Davide and Oxenl{\o}we, Leif Katsuo},
  journal={Advanced Quantum Technologies},
  volume={2},
  number={12},
  pages={1900038},
  year={2019},
  publisher={Wiley Online Library}
}

@article{mower2013high,
  title={High-dimensional quantum key distribution using dispersive optics},
  author={Mower, Jacob and Zhang, Zheshen and Desjardins, Pierre and Lee, Catherine and Shapiro, Jeffrey H and Englund, Dirk},
  journal={Physical Review A},
  volume={87},
  number={6},
  pages={062322},
  year={2013},
  publisher={APS}
}

@article{zhong2015photon,
  title={Photon-efficient quantum key distribution using time--energy entanglement with high-dimensional encoding},
  author={Zhong, Tian and Zhou, Hongchao and Horansky, Robert D and Lee, Catherine and Verma, Varun B and Lita, Adriana E and Restelli, Alessandro and Bienfang, Joshua C and Mirin, Richard P and Gerrits, Thomas and others},
  journal={New Journal of Physics},
  volume={17},
  number={2},
  pages={022002},
  year={2015},
  publisher={IOP Publishing}
}

@article{bunandar2015practical,
  title={Practical high-dimensional quantum key distribution with decoy states},
  author={Bunandar, Darius and Zhang, Zheshen and Shapiro, Jeffrey H and Englund, Dirk R},
  journal={Physical Review A},
  volume={91},
  number={2},
  pages={022336},
  year={2015},
  publisher={APS}
}

@article{lee2019large,
  title={Large-alphabet encoding for higher-rate quantum key distribution},
  author={Lee, Catherine and Bunandar, Darius and Zhang, Zheshen and Steinbrecher, Gregory R and Dixon, P Ben and Wong, Franco NC and Shapiro, Jeffrey H and Hamilton, Scott A and Englund, Dirk},
  journal={Optics express},
  volume={27},
  number={13},
  pages={17539--17549},
  year={2019},
  publisher={Optical Society of America}
}

@article{korzh2015provably,
  title={Provably secure and practical quantum key distribution over 307 km of optical fibre},
  author={Korzh, Boris and Lim, Charles Ci Wen and Houlmann, Raphael and Gisin, Nicolas and Li, Ming Jun and Nolan, Daniel and Sanguinetti, Bruno and Thew, Rob and Zbinden, Hugo},
  journal={Nature Photonics},
  volume={9},
  number={3},
  pages={163--168},
  year={2015},
  publisher={Nature Publishing Group}
}

@article{boaron2018secure,
  title={Secure quantum key distribution over 421 km of optical fiber},
  author={Boaron, Alberto and Boso, Gianluca and Rusca, Davide and Vulliez, C{\'e}dric and Autebert, Claire and Caloz, Misael and Perrenoud, Matthieu and Gras, Ga{\"e}tan and Bussi{\`e}res, F{\'e}lix and Li, Ming-Jun and others},
  journal={Physical review letters},
  volume={121},
  number={19},
  pages={190502},
  year={2018},
  publisher={APS}
}

@article{islam2017provably,
  title={Provably secure and high-rate quantum key distribution with time-bin qudits},
  author={Islam, Nurul T and Lim, Charles Ci Wen and Cahall, Clinton and Kim, Jungsang and Gauthier, Daniel J},
  journal={Science advances},
  volume={3},
  number={11},
  pages={e1701491},
  year={2017},
  publisher={American Association for the Advancement of Science}
}

@article{islam2019scalable,
  title={Scalable high-rate, high-dimensional time-bin encoding quantum key distribution},
  author={Islam, Nurul T and Lim, Charles Ci Wen and Cahall, Clinton and Qi, Bing and Kim, Jungsang and Gauthier, Daniel J},
  journal={Quantum Science and Technology},
  volume={4},
  number={3},
  pages={035008},
  year={2019},
  publisher={IOP Publishing}
}

@article{da2019experimental,
  title={Experimental demonstration of the DPTS QKD protocol over a 170 km fiber link},
  author={Da Lio, Beatrice and Bacco, Davide and Cozzolino, Daniele and Ding, Yunhong and Dalgaard, Kjeld and Rottwitt, Karsten and Oxenl{\o}we, Leif Katsuo},
  journal={Applied Physics Letters},
  volume={114},
  number={1},
  pages={011101},
  year={2019},
  publisher={AIP Publishing LLC}
}

@article{vagniluca2020efficient,
  title={Efficient time-bin encoding for practical high-dimensional quantum key distribution},
  author={Vagniluca, Ilaria and Da Lio, Beatrice and Rusca, Davide and Cozzolino, Daniele and Ding, Yunhong and Zbinden, Hugo and Zavatta, Alessandro and Oxenl{\o}we, Leif K and Bacco, Davide},
  journal={Physical Review Applied},
  volume={14},
  number={1},
  pages={014051},
  year={2020},
  publisher={APS}
}

@article{zhang2014unconditional,
  title={Unconditional security of time-energy entanglement quantum key distribution using dual-basis interferometry},
  author={Zhang, Zheshen and Mower, Jacob and Englund, Dirk and Wong, Franco NC and Shapiro, Jeffrey H},
  journal={Physical review letters},
  volume={112},
  number={12},
  pages={120506},
  year={2014},
  publisher={APS}
}

@article{brougham2013security,
  title={Security of high-dimensional quantum key distribution protocols using Franson interferometers},
  author={Brougham, Thomas and Barnett, Stephen M and McCusker, Kevin T and Kwiat, Paul G and Gauthier, Daniel J},
  journal={Journal of Physics B: Atomic, Molecular and Optical Physics},
  volume={46},
  number={10},
  pages={104010},
  year={2013},
  publisher={IOP Publishing}
}

@article{stucki2005fast,
  title={Fast and simple one-way quantum key distribution},
  author={Stucki, Damien and Brunner, Nicolas and Gisin, Nicolas and Scarani, Valerio and Zbinden, Hugo},
  journal={Applied Physics Letters},
  volume={87},
  number={19},
  pages={194108},
  year={2005},
  publisher={American Institute of Physics}
}

@article{branciard2008upper,
  title={Upper bounds for the security of two distributed-phase reference protocols of quantum cryptography},
  author={Branciard, Cyril and Gisin, Nicolas and Scarani, Valerio},
  journal={New Journal of Physics},
  volume={10},
  number={1},
  pages={013031},
  year={2008},
  publisher={IOP Publishing}
}

@inproceedings{lipinski2011white,
  title={White rabbit: A PTP application for robust sub-nanosecond synchronization},
  author={Lipi{\'n}ski, Maciej and W{\l}ostowski, Tomasz and Serrano, Javier and Alvarez, Pablo},
  booktitle={2011 IEEE International Symposium on Precision Clock Synchronization for Measurement, Control and Communication},
  pages={25--30},
  year={2011},
  organization={IEEE}
}

@article{wang2021round,
  title={Round-robin differential phase-time-shifting protocol for quantum key distribution: theory and experiment},
  author={Wang, Kai and Vagniluca, Ilaria and Zhang, Jie and Forchhammer, S{\o}ren and Zavatta, Alessandro and Christensen, Jesper B and Bacco, Davide},
  journal={Physical Review Applied},
  volume={15},
  number={4},
  pages={044017},
  year={2021},
  publisher={APS}
}

@article{doda2021quantum,
  title={Quantum key distribution overcoming extreme noise: simultaneous subspace coding using high-dimensional entanglement},
  author={Doda, Mirdit and Huber, Marcus and Murta, Gl{\'a}ucia and Pivoluska, Matej and Plesch, Martin and Vlachou, Chrysoula},
  journal={Physical Review Applied},
  volume={15},
  number={3},
  pages={034003},
  year={2021},
  publisher={APS}
}

@inproceedings{jachura2021photon,
  title={Photon-efficient quantum key distribution using multiqubit time-bin encoding},
  author={Jachura, Micha{\l} and Jarzyna, Marcin and Paw{\l}owski, Marcin and Banaszek, Konrad},
  booktitle={International Conference on Space Optics—ICSO 2020},
  volume={11852},
  pages={118525J},
  year={2021},
  organization={International Society for Optics and Photonics}
}

@article{iqbal2020high,
  title={High-Dimensional Semiquantum Cryptography},
  author={Iqbal, Hasan and Krawec, Walter O},
  journal={IEEE Transactions on Quantum Engineering},
  volume={1},
  pages={1--17},
  year={2020},
  publisher={IEEE}
}

@article{mirhosseini2015high,
  title={High-dimensional quantum cryptography with twisted light},
  author={Mirhosseini, Mohammad and Maga{\~n}a-Loaiza, Omar S and O’Sullivan, Malcolm N and Rodenburg, Brandon and Malik, Mehul and Lavery, Martin PJ and Padgett, Miles J and Gauthier, Daniel J and Boyd, Robert W},
  journal={New Journal of Physics},
  volume={17},
  number={3},
  pages={033033},
  year={2015},
  publisher={IOP Publishing}
}

@article{canas2017high,
  title={High-dimensional decoy-state quantum key distribution over multicore telecommunication fibers},
  author={Ca{\~n}as, G and Vera, N and Cari{\~n}e, J and Gonz{\'a}lez, P and Cardenas, J and Connolly, PWR and Przysiezna, A and G{\'o}mez, ES and Figueroa, M and Vallone, G and others},
  journal={Physical Review A},
  volume={96},
  number={2},
  pages={022317},
  year={2017},
  publisher={APS}
}

@article{bouchard2018experimental,
  title={Experimental investigation of high-dimensional quantum key distribution protocols with twisted photons},
  author={Bouchard, Fr{\'e}d{\'e}ric and Heshami, Khabat and England, Duncan and Fickler, Robert and Boyd, Robert W and Englert, Berthold-Georg and S{\'a}nchez-Soto, Luis L and Karimi, Ebrahim},
  journal={Quantum},
  volume={2},
  pages={111},
  year={2018},
  publisher={Verein zur F{\"o}rderung des Open Access Publizierens in den Quantenwissenschaften}
}

@article{bouchard2019quantum,
  title={Quantum process tomography of a high-dimensional quantum communication channel},
  author={Bouchard, Fr{\'e}d{\'e}ric and Hufnagel, Felix and Koutn{\`y}, Dominik and Abbas, Aazad and Sit, Alicia and Heshami, Khabat and Fickler, Robert and Karimi, Ebrahim},
  journal={Quantum},
  volume={3},
  pages={138},
  year={2019},
  publisher={Verein zur F{\"o}rderung des Open Access Publizierens in den Quantenwissenschaften}
}

@article{xu2020secure,
  title={Secure quantum key distribution with realistic devices},
  author={Xu, Feihu and Ma, Xiongfeng and Zhang, Qiang and Lo, Hoi-Kwong and Pan, Jian-Wei},
  journal={Reviews of Modern Physics},
  volume={92},
  number={2},
  pages={025002},
  year={2020},
  publisher={APS}
}

@article{bouchard2021quantum,
  title={Quantum communication with ultrafast time-bin qubits},
  author={Bouchard, Fr{\'e}d{\'e}ric and England, Duncan and Bustard, Philip J and Heshami, Khabat and Sussman, Benjamin},
  journal={arXiv preprint arXiv:2106.09833},
  year={2021}
}

@article{trenyi2021zero,
  title={Zero-error attack against coherent-one-way quantum key distribution},
  author={Tr{\'e}nyi, R{\'o}bert and Curty, Marcos},
  journal={New Journal of Physics},
  volume={23},
  number={9},
  pages={093005},
  year={2021},
  publisher={IOP Publishing}
}

@article{curty2021foiling,
  title={Foiling zero-error attacks against coherent-one-way quantum key distribution},
  author={Curty, Marcos},
  journal={Physical Review A},
  volume={104},
  number={6},
  pages={062417},
  year={2021},
  publisher={APS}
}

@article{gisin2004towards,
  title={Towards practical and fast quantum cryptography},
  author={Gisin, Nicolas and Ribordy, Gr{\'e}goire and Zbinden, Hugo and Stucki, Damien and Brunner, Nicolas and Scarani, Valerio},
  journal={arXiv preprint quant-ph/0411022},
  year={2004}
}

@article{stucki2009high,
  title={High rate, long-distance quantum key distribution over 250 km of ultra low loss fibres},
  author={Stucki, Damien and Walenta, Nino and Vannel, Fabien and Thew, Robert Thomas and Gisin, Nicolas and Zbinden, Hugo and Gray, S and Towery, CR and Ten, S},
  journal={New Journal of Physics},
  volume={11},
  number={7},
  pages={075003},
  year={2009},
  publisher={IOP Publishing}
}

@article{stucki2009continuous,
  title={Continuous high speed coherent one-way quantum key distribution},
  author={Stucki, Damien and Barreiro, Claudio and Fasel, Sylvain and Gautier, Jean-Daniel and Gay, Olivier and Gisin, Nicolas and Thew, Rob and Thoma, Yann and Trinkler, Patrick and Vannel, Fabien and others},
  journal={Optics express},
  volume={17},
  number={16},
  pages={13326--13334},
  year={2009},
  publisher={Optical Society of America}
}

@article{walenta2014fast,
  title={A fast and versatile quantum key distribution system with hardware key distillation and wavelength multiplexing},
  author={Walenta, Nino and Burg, Andreas and Caselunghe, Dario and Constantin, Jeremy and Gisin, Nicolas and Guinnard, Olivier and Houlmann, Rapha{\"e}l and Junod, Pascal and Korzh, Boris and Kulesza, Natalia and others},
  journal={New Journal of Physics},
  volume={16},
  number={1},
  pages={013047},
  year={2014},
  publisher={IOP Publishing}
}

@article{sibson2017chip,
  title={Chip-based quantum key distribution},
  author={Sibson, Philip and Erven, Chris and Godfrey, Mark and Miki, Shigehito and Yamashita, Taro and Fujiwara, Mikio and Sasaki, Masahide and Terai, Hirotaka and Tanner, Michael G and Natarajan, Chandra M and others},
  journal={Nature communications},
  volume={8},
  number={1},
  pages={1--6},
  year={2017},
  publisher={Nature Publishing Group}
}

@article{roberts2017modulator,
  title={Modulator-Free Coherent-One-Way Quantum Key Distribution},
  author={Roberts, George L and Lucamarini, Marco and Dynes, James F and Savory, Seb J and Yuan, ZL and Shields, Andrew J},
  journal={Laser \& Photonics Reviews},
  volume={11},
  number={4},
  pages={1700067},
  year={2017},
  publisher={Wiley Online Library}
}

@article{sibson2017integrated,
  title={Integrated silicon photonics for high-speed quantum key distribution},
  author={Sibson, Philip and Kennard, Jake E and Stanisic, Stasja and Erven, Chris and O’Brien, Jeremy L and Thompson, Mark G},
  journal={Optica},
  volume={4},
  number={2},
  pages={172--177},
  year={2017},
  publisher={Optical Society of America}
}

@article{dai2020pass,
  title={Pass-block architecture for distributed-phase-reference quantum key distribution using silicon photonics},
  author={Dai, Jincheng and Zhang, Lei and Fu, Xin and Zheng, Xuezhe and Yang, Lin},
  journal={Optics Letters},
  volume={45},
  number={7},
  pages={2014--2017},
  year={2020},
  publisher={Optical Society of America}
}

@article{moroder2012security,
  title={Security of distributed-phase-reference quantum key distribution},
  author={Moroder, Tobias and Curty, Marcos and Lim, Charles Ci Wen and Zbinden, Hugo and Gisin, Nicolas and others},
  journal={Physical review letters},
  volume={109},
  number={26},
  pages={260501},
  year={2012},
  publisher={APS}
}

@article{branciard2006zero,
  title={Zero-error attacks and detection statistics in the coherent one-way protocol for quantum cryptography},
  author={Branciard, Cyril and Gisin, Nicolas and Lutkenhaus, Norbert and Scarani, Valerio},
  journal={arXiv preprint quant-ph/0609090},
  year={2006}
}

@article{renner2008security,
  title={Security of quantum key distribution},
  author={Renner, Renato},
  journal={International Journal of Quantum Information},
  volume={6},
  number={01},
  pages={1--127},
  year={2008},
  publisher={World Scientific}
}

@article{tomamichel2011uncertainty,
  title={Uncertainty relation for smooth entropies},
  author={Tomamichel, Marco and Renner, Renato},
  journal={Physical review letters},
  volume={106},
  number={11},
  pages={110506},
  year={2011},
  publisher={APS}
}

@article{lofberg2004proceedings,
  title={Proceedings of the CACSD Conference},
  author={L{\"o}fberg, J},
  year={2004},
  publisher={IEEE Piscataway}
}

@incollection{andersen2000mosek,
  title={The MOSEK interior point optimizer for linear programming: an implementation of the homogeneous algorithm},
  author={Andersen, Erling D and Andersen, Knud D},
  booktitle={High performance optimization},
  pages={197--232},
  year={2000},
  publisher={Springer}
}

@article{lib2022processing,
  title={Processing Entangled Photons in High Dimensions with a Programmable Light Converter},
  author={Lib, Ohad and Sulimany, Kfir and Bromberg, Yaron},
  journal={Physical Review Applied},
  volume={18},
  number={1},
  pages={014063},
  year={2022},
  publisher={APS}
}

@article{sulimany2022all,
  title={All-fiber source and sorter for multimode correlated photons},
  author={Sulimany, Kfir and Bromberg, Yaron},
  journal={npj Quantum Information},
  volume={8},
  number={1},
  pages={1--5},
  year={2022},
  publisher={Nature Publishing Group}
}

@article{devetak2005distillation,
  title={Distillation of secret key and entanglement from quantum states},
  author={Devetak, Igor and Winter, Andreas},
  journal={Proceedings of the Royal Society A: Mathematical, Physical and engineering sciences},
  volume={461},
  number={2053},
  pages={207--235},
  year={2005},
  publisher={The Royal Society}
}

@article{ali2007large,
  title={Large-alphabet quantum key distribution using energy-time entangled bipartite states},
  author={Ali-Khan, Irfan and Broadbent, Curtis J and Howell, John C},
  journal={Physical review letters},
  volume={98},
  number={6},
  pages={060503},
  year={2007},
  publisher={APS}
}

@article{cabello2000quantum,
  title={Quantum key distribution in the Holevo limit},
  author={Cabello, Ad{\'a}n},
  journal={Physical Review Letters},
  volume={85},
  number={26},
  pages={5635},
  year={2000},
  publisher={APS}
}

@article{lo2014secure,
  title={Secure quantum key distribution},
  author={Lo, Hoi-Kwong and Curty, Marcos and Tamaki, Kiyoshi},
  journal={Nature Photonics},
  volume={8},
  number={8},
  pages={595--604},
  year={2014},
  publisher={Nature Publishing Group UK London}
}

@article{bulla2023nonlocal,
  title={Nonlocal Temporal Interferometry for Highly Resilient Free-Space Quantum Communication},
  author={Bulla, Lukas and Pivoluska, Matej and Hjorth, Kristian and Kohout, Oskar and Lang, Jan and Ecker, Sebastian and Neumann, Sebastian P and Bittermann, Julius and Kindler, Robert and Huber, Marcus and others},
  journal={Physical Review X},
  volume={13},
  number={2},
  pages={021001},
  year={2023},
  publisher={APS}
}

@article{reimer2016generation,
  title={Generation of multiphoton entangled quantum states by means of integrated frequency combs},
  author={Reimer, Christian and Kues, Michael and Roztocki, Piotr and Wetzel, Benjamin and Grazioso, Fabio and Little, Brent E and Chu, Sai T and Johnston, Tudor and Bromberg, Yaron and Caspani, Lucia and others},
  journal={Science},
  volume={351},
  number={6278},
  pages={1176--1180},
  year={2016},
  publisher={American Association for the Advancement of Science}
}

@article{kues2019quantum,
  title={Quantum optical microcombs},
  author={Kues, Michael and Reimer, Christian and Lukens, Joseph M and Munro, William J and Weiner, Andrew M and Moss, David J and Morandotti, Roberto},
  journal={Nature Photonics},
  volume={13},
  number={3},
  pages={170--179},
  year={2019},
  publisher={Nature Publishing Group UK London}
}

@article{winick2018reliable,
  title={Reliable numerical key rates for quantum key distribution},
  author={Winick, Adam and L{\"u}tkenhaus, Norbert and Coles, Patrick J},
  journal={Quantum},
  volume={2},
  pages={77},
  year={2018},
  publisher={Verein zur F{\"o}rderung des Open Access Publizierens in den Quantenwissenschaften}
}

@article{hu2022robust,
  title={Robust interior point method for quantum key distribution rate computation},
  author={Hu, Hao and Im, Jiyoung and Lin, Jie and L{\"u}tkenhaus, Norbert and Wolkowicz, Henry},
  journal={Quantum},
  volume={6},
  pages={792},
  year={2022},
  publisher={Verein zur F{\"o}rderung des Open Access Publizierens in den Quantenwissenschaften}
}

@article{araujo2023quantum,
  title={Quantum key distribution rates from semidefinite programming},
  author={Ara{\'u}jo, Mateus and Huber, Marcus and Navascu{\'e}s, Miguel and Pivoluska, Matej and Tavakoli, Armin},
  journal={Quantum},
  volume={7},
  pages={1019},
  year={2023},
  publisher={Verein zur F{\"o}rderung des Open Access Publizierens in den Quantenwissenschaften}
}

@article{ecker2019overcoming,
  title={Overcoming noise in entanglement distribution},
  author={Ecker, Sebastian and Bouchard, Fr{\'e}d{\'e}ric and Bulla, Lukas and Brandt, Florian and Kohout, Oskar and Steinlechner, Fabian and Fickler, Robert and Malik, Mehul and Guryanova, Yelena and Ursin, Rupert and others},
  journal={Physical Review X},
  volume={9},
  number={4},
  pages={041042},
  year={2019},
  publisher={APS}
}

@article{bunandar2018metropolitan,
  title={Metropolitan quantum key distribution with silicon photonics},
  author={Bunandar, Darius and Lentine, Anthony and Lee, Catherine and Cai, Hong and Long, Christopher M and Boynton, Nicholas and Martinez, Nicholas and DeRose, Christopher and Chen, Changchen and Grein, Matthew and others},
  journal={Physical Review X},
  volume={8},
  number={2},
  pages={021009},
  year={2018},
  publisher={APS}
}

@article{thew2004bell,
  title={Bell-type test of energy-time entangled qutrits},
  author={Thew, Robert Thomas and Acin, Antonio and Zbinden, Hugo and Gisin, Nicolas},
  journal={Physical review letters},
  volume={93},
  number={1},
  pages={010503},
  year={2004},
  publisher={APS}
}

@article{brendel1999pulsed,
  title={Pulsed energy-time entangled twin-photon source for quantum communication},
  author={Brendel, J{\"u}rgen and Gisin, Nicolas and Tittel, Wolfgang and Zbinden, Hugo},
  journal={Physical Review Letters},
  volume={82},
  number={12},
  pages={2594},
  year={1999},
  publisher={APS}
}

@article{jha2008exploring,
  title={Exploring energy-time entanglement using geometric phase},
  author={Jha, Anand Kumar and Malik, Mehul and Boyd, Robert W},
  journal={Physical review letters},
  volume={101},
  number={18},
  pages={180405},
  year={2008},
  publisher={APS}
}

@article{maclean2018direct,
  title={Direct characterization of ultrafast energy-time entangled photon pairs},
  author={MacLean, Jean-Philippe W and Donohue, John M and Resch, Kevin J},
  journal={Physical review letters},
  volume={120},
  number={5},
  pages={053601},
  year={2018},
  publisher={APS}
}

@article{vaziri2002experimental,
  title={Experimental two-photon, three-dimensional entanglement for quantum communication},
  author={Vaziri, Alipasha and Weihs, Gregor and Zeilinger, Anton},
  journal={Physical review letters},
  volume={89},
  number={24},
  pages={240401},
  year={2002},
  publisher={APS}
}

@article{leach2010quantum,
  title={Quantum correlations in optical angle--orbital angular momentum variables},
  author={Leach, Jonathan and Jack, Barry and Romero, Jacqui and Jha, Anand K and Yao, Alison M and Franke-Arnold, Sonja and Ireland, David G and Boyd, Robert W and Barnett, Stephen M and Padgett, Miles J},
  journal={Science},
  volume={329},
  number={5992},
  pages={662--665},
  year={2010},
  publisher={American Association for the Advancement of Science}
}

@article{erhard2018twisted,
  title={Twisted photons: new quantum perspectives in high dimensions},
  author={Erhard, Manuel and Fickler, Robert and Krenn, Mario and Zeilinger, Anton},
  journal={Light: Science \& Applications},
  volume={7},
  number={3},
  pages={17146--17146},
  year={2018},
  publisher={Nature Publishing Group}
}

@article{howell2004realization,
  title={Realization of the Einstein-Podolsky-Rosen paradox using momentum-and position-entangled photons from spontaneous parametric down conversion},
  author={Howell, John C and Bennink, Ryan S and Bentley, Sean J and Boyd, Robert W},
  journal={Physical review letters},
  volume={92},
  number={21},
  pages={210403},
  year={2004},
  publisher={APS}
}

@article{schaeff2015experimental,
  title={Experimental access to higher-dimensional entangled quantum systems using integrated optics},
  author={Schaeff, Christoph and Polster, Robert and Huber, Marcus and Ramelow, Sven and Zeilinger, Anton},
  journal={Optica},
  volume={2},
  number={6},
  pages={523--529},
  year={2015},
  publisher={Optica Publishing Group}
}

@article{wang2018multidimensional,
  title={Multidimensional quantum entanglement with large-scale integrated optics},
  author={Wang, Jianwei and Paesani, Stefano and Ding, Yunhong and Santagati, Raffaele and Skrzypczyk, Paul and Salavrakos, Alexia and Tura, Jordi and Augusiak, Remigiusz and Man{\v{c}}inska, Laura and Bacco, Davide and others},
  journal={Science},
  volume={360},
  number={6386},
  pages={285--291},
  year={2018},
  publisher={American Association for the Advancement of Science}
}

@article{srivastav2022quick,
  title={Quick Quantum Steering: Overcoming Loss and Noise with Qudits},
  author={Srivastav, Vatshal and Valencia, Natalia Herrera and McCutcheon, Will and Leedumrongwatthanakun, Saroch and Designolle, S{\'e}bastien and Uola, Roope and Brunner, Nicolas and Malik, Mehul},
  journal={Physical Review X},
  volume={12},
  number={4},
  pages={041023},
  year={2022},
  publisher={APS}
}

@article{sit2017high,
  title={High-dimensional intracity quantum cryptography with structured photons},
  author={Sit, Alicia and Bouchard, Fr{\'e}d{\'e}ric and Fickler, Robert and Gagnon-Bischoff, J{\'e}r{\'e}mie and Larocque, Hugo and Heshami, Khabat and Elser, Dominique and Peuntinger, Christian and G{\"u}nthner, Kevin and Heim, Bettina and others},
  journal={Optica},
  volume={4},
  number={9},
  pages={1006--1010},
  year={2017},
  publisher={Optica Publishing Group}
}

@misc{HDQKD2023,
  author = {Sulimany, Kfir and Pelc, Guy and Dudkiewicz, Rom and Korenblit, Simcha and S. Eisenberg, Hagai and Bromberg, Yaron and Ben-Or, Michael},
  title = {High dimensional coherent one way quantum key distribution},
  year = {2023},
  publisher = {GitHub},
  journal = {GitHub repository},
  howpublished = {\url{https://github.cs.huji.ac.il/guy-pelc/HD_COW_QKD}},
}

\clearpage

\maketitle

\onecolumn
\section{Supplementary Information}
\subsection{Detailed security analysis} \label{Sup_Detailed security analysis}

In this part, we will revisit the analysis done for Eve's extractable information in more detail.

\subsubsection{Eve's action and constraints}
As a reminder, we can look at Eve's action as a linear transformation \cite{branciard2008upper}
\begin{equation} \label{eq15}
|0\rangle_{A}|\varepsilon\rangle_{E} \to |0\rangle _{B} |v_{0}\rangle _{E} + \sqrt{Q \mu t}|1\rangle _{B} |p_{0}\rangle _{E}
\end{equation}

\begin{equation} \label{eq16}
|\sqrt{\mu}\rangle_{A} |\varepsilon\rangle_{E} \to |0\rangle _{B} |v_{\mu}\rangle _{E} + \sqrt{(1-(d-1)Q)\mu t}|1\rangle _{B} |p_{\mu}\rangle _{E}
\end{equation}
Where Q is the quantum bit error rate (QBER) i.e. the probability that Bob accepts the wrong bit value. $\mu$ and $t$ are the pulse occupation and the link transmission. These amplitudes of the states are chosen such that Eve does not change the QBER of the data line.

The loss of coherence is monitored by analyzing the detection events in the monitoring line. The phase between the two arms of the interferometer in the monitoring line is chosen such that two successive non-empty pulses sent by Alice will interfere destructively in one output port and constructively in the other port. We label the probability of detecting a photon at the constructive port by $P(D_c)$ and the probability of detecting a photon at the destructive port by $P(D_d)$. The loss of coherence by Eve's attack is measured by the visibility:

\begin{equation} \label{eq17}
V = \frac{P(D_c)-P(D_d)}{P(D_c)+P(D_d)}
\end{equation}

Assuming $\mu t \ll 1$ we can neglect two-photon terms. Eve's action on a successive  pair of occupied pulses $|\sqrt{\mu}\rangle|\sqrt{\mu}\rangle$ is then given by:

\begin{equation} \label{eq18}
\begin{gathered} 
|\sqrt{\mu},\sqrt{\mu}\rangle_A |\varepsilon,\varepsilon\rangle_{E} \to
|0,0\rangle_B |v_\mu ,v_\mu \rangle _E 
+ \sqrt{(1-(d-1)Q)\mu t}[|1,0\rangle_B |p_\mu ,v_\mu \rangle _E  +|0,1\rangle_B |v_\mu ,p_\mu \rangle _E]
\end{gathered}
\end{equation}

The action of the interferometer at Bob's end yields:

\begin{equation} \label{eq19}
\begin{gathered} 
|1,0\rangle_B |p_\mu ,v_\mu \rangle _E  +|0,1\rangle_B |v_\mu ,p_\mu \rangle _E \to 
|D_c\rangle_B(|p_\mu ,v_\mu \rangle _E + |v_\mu ,p_\mu \rangle _E)+ |D_d\rangle_B(|p_\mu ,v_\mu \rangle _E - |v_\mu ,p_\mu \rangle _E)
\end{gathered}
\end{equation}
where we define the constructive and destructive output modes of the interferometer by $|D_{c/d}\rangle_B = \frac{1}{\sqrt{2}} (|1,0\rangle_B\pm{}|0,1\rangle_B)$. The probability that the photon is detected at the constructive/destructive detector is thus given by $P(D_{c/d}) \propto |(|p_\mu ,v_\mu \rangle _E \pm{} |v_\mu ,p_\mu \rangle _E)|^2$. The visibility constraint on Eve's action is therefore given by: 

\begin{equation} \label{eq20}
\begin{gathered} 
V = |\langle p_{\mu}|v_{\mu}\rangle|^{2}
\end{gathered}
\end{equation}

The third constraint on Eve's transformation is that it must be unitary. In the $\mu t \ll 1$ limit we get 

\begin{equation} \label{eq21}
\begin{gathered} 
\langle v_0 | v_\mu \rangle = e^{- \mu /2} 
\end{gathered}
\end{equation}

To compute the amount of information that can be extracted by Eve, quantified by the Holevo information, we first need to analyze her action on a qudit with occupation $\mu^{(i)}$ in the $i^{th}$ time bin. Neglecting all multiple photon terms, Eve's action can be presented by:
\begin{equation} \label{eq22}
\begin{gathered}
|0,...,0,\sqrt{\mu}^{(i)},0,...,0\rangle_A|\varepsilon,...,\varepsilon\rangle_E \to \\
|0,...,0\rangle_B \otimes |V_{i}\rangle 
+  \sqrt{(1-(d-1)Q)\mu t}|0,...,0,1^{(i)},0...,0\rangle_B \otimes |C_{i}\rangle 
+ \sum_{k=1...d, k\neq i}\sqrt{Q \mu t} |0,...,0,1^{(k)},0...,0\rangle_B |W_{i,k}\rangle,
\end{gathered}
\end{equation} 
where $|V_{i}\rangle = |v_{0},...,v_{0},v_{\mu}^{(i)},v_{0},...,v_{0}\rangle_E$ is Eve's state representing the case where she sends a vacuum state at time bin $i$, $|C_{i}\rangle = |v_{0},...,v_{0},p_{\mu}^{(i)},v_{0},...,v_{0}\rangle_E$ is Eve's state representing the case where she sends to Bob a photon at the right time bin $i$, and $|W_{i,k}\rangle = |v_{0},...,v_{0},p_{0}^{(k)},v_{0},...,v_{0},v_{\mu}^{(i)},v_{0},...,v_{0}\rangle_E$ is Eve's state representing the case where she sends to Bob a photon at the wrong time bin $k$. 

The density matrix representing Eve's subsystem, conditioned on the event where Alice transmits a pulse at time bin $i$ and Bob detects a photon at any arbitrary time bin, can be expressed as follows: 
\begin{equation} \label{eq23}
\begin{gathered}
\rho_{E}^{A=i} = (1-(d-1)Q)|C_{i}\rangle \langle C_{i}| 
+ \sum_{k=1...d, k\neq i}Q |W_{i,k}\rangle \langle W_{i,k}| 
\end{gathered}
\end{equation}
The density matrix of Eve's subsystem is conditioned by the event where Bob detects a pulse at time bin $i$ and Alice sends the pulse at an arbitrary time bin:
\begin{equation} \label{eq24}
\begin{gathered}
\rho_{E}^{B=i} = (1-(d-1)Q)|C_{i}\rangle \langle C_{i}| 
+ \sum_{k=1...d, k\neq i}Q |W_{k,i}\rangle \langle W_{k,i}|
\end{gathered}
\end{equation}

\subsubsection{Eve's Holevo information}

Eve's information is bounded by the Holevo bound on the Alice-Eve channel ($\chi_{AE}$) and on the Bob-Eve channel ($\chi_BE$). The maximum amount of information Eve can extract is given by $\max\{\chi_{BE},\chi_{AE}\}$. We start by analyzing $\chi_{BE}$, and we will then show that $\chi_{BE}>\chi_{AE}$.

As explained in the main text, Eve can choose the $|p_0\rangle_E$ state to be orthogonal to all other states. We now use this to simplify Eve's Holevo information. We start by choosing an orthonormal set of states $G'$ that span the space that describes Eve's system for a single time bin, where $|p_0\rangle_E\in G'$ is one of the states in the set. The basis for Eve's system that spans a $d$-dimension qudit is simply the tensor product of this basis, i.e. $|v_{1},...,v_{d}\rangle_E \in G = |v_{1}\rangle_E \otimes ... \otimes |v_{d}\rangle_E \in \prod_{i=1}^{d}G' $. 

We can now split the space $G$ to d+2 groups: $G_{i}, i=1...d$ will consist of states with $|p_{0}\rangle_E$ in the $i^{th}$ time bin, and other vectors from $G'$ in rest of the time bins. $G_{d+1}$ will consist states without $|p_{0}\rangle_E$ at any time bin, and $G_{d+2}$ will consist states $|p_{0}\rangle_E$ in more than one time bin. We further define $P_{l}=\sum_{|v\rangle \in G_{l} }|v\rangle\langle v|$ and since $\sum P_{l} = I$, we can express $\chi_{BE}$ by:

\begin{equation} \label{eq25}
\begin{gathered}
\chi_{BE}  = S\left(\sum_{i=1}^{d} \frac{1}{d}\rho_{E}^{B=i}\right) - \sum_{i=1}^{d}\frac{1}{d}S\left( \rho_{E}^{B=i}\right)  \\ 
= S\left(\sum_{j=1}^{d+2}P_{j}\sum_{i=1...d} \frac{1}{d}\rho_{E}^{B=i}\right) - \sum_{i=1...d}\frac{1}{d}S\left(\sum_{j=1}^{d+2}P_{j} \rho_{E}^{B=i}\right)\\
= S\left(\sum_{j=1}^{d+2}P_{j}\sum_{i=1}^{d} \frac{1}{d}\left( (1-(d-1)Q)|C_{i}\rangle \langle C_{i}|
+ \sum_{k=1, k\neq i}^{d}Q |W_{k,i}\rangle \langle W_{k,i}|\right)\right) \\
- \sum_{i=1}^{d}\frac{1}{d}S\left(\sum_{j=1}^{d+2}P_{j} \left( (1-(d-1)Q)|C_{i}\rangle \langle C_{i}| 
+ \sum_{k=1, k\neq i}^{d}Q |W_{k,i}\rangle \langle W_{k,i}|\right)\right) 
\end{gathered}
\end{equation}

We can simplify this expression using the fact that the von-Neumann entropy of a block-diagonal matrix equals the sum of the entropies of the blocks along the diagonal. We claim that the matrices that appear in eq.(\ref{eq26}) are all block-diagonal, by the construction of the sets $G_i$. To show this, it is enough to prove that $_E\langle v|\rho_{E}^{A=k}|u\rangle_E = 0$ for every $k$ and every $|v\rangle_E\in G_i$, $|u\rangle_E \in G_j, j\neq i$. Since each density matrix is by itself a sum of density matrices of pure states, we are left with proving that $_E\langle v|\varphi \rangle \langle \varphi |u\rangle_E = 0$ for every state $|\varphi\rangle_E$ describing Eve's system after a qudit was sent by Alice and received by Bob, represented by $|C_{i}\rangle=|v_{0},...,v_{0},p_{\mu}^{(i)},v_{0},...,v_{0}\rangle_E, i\in {1...d}$ or  
$|W_{i,k}\rangle = |v_{0},...,v_{0},p_{0}^{(k)},v_{0},...,v_{0},v_{\mu}^{(i)},v_{0},...,v_{0}\rangle_E,\ i\neq k\in {1...d}$. It is easy to verify by inspection that $ _E\langle v|C_{i}\rangle \langle C_{i}|u\rangle_E = 0$ if $|v\rangle_E$ or $|u\rangle_E$ are not both in $G_{d+1}$. It can also be verified that $ _E\langle v|W_{i,k}\rangle \langle W_{i,k}|u\rangle_E = 0$ if $|v\rangle_E$ or $|u\rangle_E$ are not both in $G_{k}$. In other words, if $|u\rangle_E$ and $|v\rangle_E$ are contained in different sets, then $_E\langle v|W_{i,k}\rangle \langle W_{i,k}|u\rangle_E = 0$. We can therefore move the sum over $j$ in eq.(\ref{eq26}) outside the entropy $S$:

\begin{equation} \label{eq26}
\begin{gathered}
\chi_{BE}  =   \sum_{j=1}^{d+2}S\left(P_{j}\sum_{i=1}^{d} \frac{1}{d}\left( (1-(d-1)Q)|C_{i}\rangle \langle C_{i}|
+ \sum_{k=1, k\neq i}^{d}Q |W_{k,i}\rangle \langle W_{k,i}|\right)\right) \\
- \sum_{i=1}^{d}\frac{1}{d}\sum_{j=1}^{d+2}S\left(P_{j} \left( (1-(d-1)Q)|C_{i}\rangle \langle C_{i}| 
+ \sum_{k=1, k\neq i}^{d}Q |W_{k,i}\rangle \langle W_{k,i}|\right)\right) 
\end{gathered}
\end{equation}

Since $P_{l}|C_{j}\rangle=0$ if $l\neq d+1$ and $P_{l}|W_{i,k}\rangle = 0$ if $l\neq k$, out of all the terms in the sum over $j$ in eq.(\ref{eq26}) we are left with the terms $P_{d+1}|C_{i}\rangle = |C_{i}\rangle$ and $P_{k} |W_{k,i}\rangle=|W_{k,i}\rangle$, which greatly simplifies the expression for the Holevo bound: 

\begin{equation} \label{eq27}
\begin{gathered}
\chi_{BE}  = S\left(\sum_{i=1}^{d} \frac{ (1-(d-1)Q)}{d}|C_{i}\rangle \langle C_{i}|)\right)
+ \sum_{k=1}^{d}\frac{1}{d} S\left( \sum_{i=1,i\neq k}^{d} Q |W_{k,i}\rangle \langle W_{k,i}|\right) 
\\
- \sum_{i=1}^{d}\frac{1}{d}S\left( (1-(d-1)Q)|C_{i}\rangle \langle C_{i}|\right)- \sum_{i=1}^{d}\frac{1}{d}S\left(\sum_{k=1, k\neq i}^{d}Q |W_{k,i}\rangle \langle W_{k,i}|\right)
\end{gathered}
\end{equation}

We define $d-1$ vectors $|W'_{i}\rangle=|v_{0},...,v_{0},v_{\mu}^{(i)},v_{0},...,v_{0}\rangle_E, i\in {1...d-1}$ such that $|W'_{i}\rangle \otimes |p_0\rangle = |W_{i,d}\rangle$ and $|W_{i,j}\rangle$ are equivalent up to reordering the order of the time bins. Since the entropy is additive for independent systems, rearranging the order of the terms in eq.(\ref{eq27}), we get:

\begin{equation} \label{eq28}
\begin{gathered}
\chi_{BE}  =
\sum_{k=1}^{d} \left[S\left(\sum_{i=1}^{d-1} \frac{Q}{d}  |W'_{i}\rangle \langle W'_{i}| \right)\right. 
\left. -\frac{1}{d} S \left(\sum_{i=1}^{d-1}Q|W'_{i}\rangle \langle W'_{i}| \right)
\right] \\ 
+ S\left(\sum_{i=1}^{d} \frac{1-(d-1)Q}{d} |C_{i}\rangle \langle C_{i}| \right)
-\frac{1}{d}\sum_{i=1}^{d} S \left(\left(1-(d-1)Q\right)|C_{i}\rangle \langle C_{i}| \right) 
\end{gathered}
\end{equation}

Repeating the above steps from the Holevo bound on the channel between Alice and Eve yields:
\begin{equation} \label{eq29}
\begin{gathered}
\chi_{AE}  =
\sum_{k=1}^{d} \left[S\left(\sum_{i=1}^{d-1} \frac{Q}{d}  |W'_{i}\rangle \langle W'_{i}| \right)\right. 
\left. -\frac{1}{d}\sum_{i=1}^{d-1} S \left(Q|W'_{i}\rangle \langle W'_{i}| \right)
\right] \\ 
+ S\left(\sum_{i=1}^{d-1} \frac{1-(d-1)Q}{d} |C_{i}\rangle \langle C_{i}| \right)
-\frac{1}{d}\sum_{i=1}^{d} S \left(\left(1-(d-1)Q\right)|C_{i}\rangle \langle C_{i}| \right) 
\end{gathered}
\end{equation}

To show that the Holevo of the channel between Eve and Bob is higher than the Holveo of Eve and Alics, we notice that $\chi_{BE}-\chi_{AE} = \sum_{i=1}^{d-1} S \left(Q|W'_{i}\rangle \langle W'_{i}| \right) - S \left(\sum_{i=1}^{d-1}Q|W'_{i}\rangle \langle W'_{i}| \right)=(d-1)S(Q) - S\left(\sum_{i=1}^{d-1}Q|W'_{i}\rangle \langle W'_{i}| \right)$. The dimension of the matrix in the second term is at most $d-1$ and thus it cannot have more than $d-1$ nonzero eigenvalues. The maximal entropy of the second term is achieved when all of the eigenvalues are equal, and since its trace is $(d-1)Q$, we conclude that $\sum_{i=1}^{d-1}Q|W'_{i}\rangle \langle W'_{i}| \leq(d-1)S(Q)$. 
This can also be seen intuitively, as Eve's information over Alice or Bob's state is the same when the correct qudit was transferred, but when an error was passed Eve knows for sure what Bob got but has only partial certainty over what Alice sent. This yields that the Holevo-information will be maximal with Bob.

To calculate the entropy of the above matrices we need to find their eigenvalues. We notice that both $|W'_{i}\rangle$ and $|C_{i}\rangle$ have the same form, $|u,...u,v,u,...,u\rangle$ for some dimension ($d$ or $d-1$). We now find in general the eigenvalues of a matrix $M=\sum_{i=1...n} \overbrace{ |u,...u,v^{(i)},u,...,u\rangle}^{n\ terms} \langle u,...u,v^{(i)},u,...,u|$. We can view $|v\rangle$ as $|v\rangle=\alpha |u\rangle+\beta |u^{\perp}\rangle \ s.t.\ \langle u|u^{\perp}\rangle=0$ and $|u^{\perp}\rangle$ is a unit vector, which gives us $|\langle v|v\rangle|^2=|\alpha|^{2}+|\beta|^{2}=1$ and $\alpha = \langle u|v\rangle$. Now we can define the vectors $|U\rangle=|u,...u\rangle$ and $|V_{i}\rangle=|u,...u,u^{\perp (i)},u,...,u\rangle$ all orthogonal to each other, and get from linearity $M = \sum_{i=1...n}(\alpha |U\rangle + \beta |V_{i}\rangle)(\alpha^* \langle U| + \beta^* \langle V_{i}|)=n|\alpha|^{2}|U\rangle\langle U| + \sum_{i=1...n}\left(\alpha\beta^*|U\rangle\langle V_{i}|+\alpha^*\beta|V_{i}\rangle\langle U|+|\beta|^{2}|V_{i}\rangle\langle V_{i}|\right)$. By narrowing the matrix to the space spanned by $|U\rangle$ and $|V_{i}\rangle$, we get:

\begin{equation} \label{eq30}
M=\left[\begin{array}{ccccc}
n|\alpha|^{2}&\alpha\beta^*&\alpha\beta^* & \cdots & \alpha\beta^*\\
\alpha^*\beta&|\beta|^{2}& 0 & \cdots &0\\
\alpha^*\beta&0&|\beta|^{2} & \cdots &0\\
\vdots &\vdots &\vdots &\ddots &\vdots\\
\alpha^*\beta & 0 & \cdots &0& |\beta|^{2}\\
\end{array}\right]
 = |\beta|^{2}I + \left[\begin{array}{cccc}
n|\alpha|^{2}-|\beta|^{2}&\alpha\beta^*&\cdots & \alpha\beta^*\\
\alpha^*\beta& 0 & \cdots &0\\
\vdots &&\ddots &\vdots\\
\alpha^*\beta & 0 & \cdots &0\\
\end{array}\right] = |\beta|^{2}I + M'
\end{equation}
To obtain the eigenvalues $M$ we can find the roots of the characteristic polynomial of $M'$ and add $|\beta|^{2}$ to all of them. $|\lambda I-M'| = (\lambda-n|\alpha|^{2}+|\beta|^{2})\lambda^{n} - n|\alpha|^{2}|\beta|^{2}\lambda^{n-1}= \lambda^{n-1}(\lambda+|\beta|^{2})(\lambda-n|\alpha|^{2})$, so the eigenvalues of $M'$ are $n|\alpha|^{2}, -|\beta|^{2},$ and $0$ with multiplicity $n-1$. Therefore the eigenvalues of $M$ are $n|\alpha|^{2} + |\beta|^{2},0,$ and $|\beta|^{2}$ with multiplicity $n-1$, where the zero eigenvalue does not affect the entropy.

Substituting the above eigenvalues into the expression for Holevo bound on Bob-Eve channel eq.(\ref{eq28}) information yields:

\begin{equation} \label{eq31}
\begin{gathered}
\chi_{BE}  =
\sum_{k=1}^{d} \left[\left((d-2)S(\frac{Q}{d}(1- |\langle v_0|v_{\mu}\rangle|^{2}))\right) + S(\frac{Q}{d}(1+ (d-2)|\langle v_0|v_{\mu}\rangle|^{2}))\right.\\
\left.-\frac{d-2}{d}\left(S(Q(1- |\langle v_0|v_{\mu}\rangle|^{2}))\right) -\frac{1}{d} S(Q(1+ (d-2)|\langle v_0|v_{\mu}\rangle|^{2}))\right] \\ 
+ (d-1)S\left(\frac{1-(d-1)Q}{d}(1-|\langle v_0|p_{\mu}\rangle|^{2}) \right) + S\left(\frac{1-(d-1)Q}{d}(1+(d-1)|\langle v_0|p_{\mu}\rangle|^{2})  \right)
-\frac{1}{d}\sum_{i=1}^{d} S \left(\left(1-(d-1)Q\right)\right) 
\end{gathered}
\end{equation}

Imposing the unitary constraint for $\langle v_0|v_{\mu}\rangle = e^{-\mu/2}$, and  noticing that the first line in eq.(\ref{eq31}) can be simplified by $S(\frac{1}{d}A)-\frac{1}{d}S(A) = \sum-\frac{\lambda_i}{d} \log_{2}(\frac{\lambda_i}{d}) - \frac{1}{d}\sum-\lambda_i \log_{2}(\lambda_i) = S(\frac{1}{d})\sum\lambda_i$, yields the following expression for the Holevo bound: 

\begin{equation} \label{eq32} 
\begin{gathered}
\chi_{BE}  = Q(d-1)\log_{2}(d) \\
+ S\left( \frac{1-(d-1)Q}{d} \left((d-1)|\langle v_0 | p_\mu \rangle| ^{2} +1 \right) \right)
+ (d-1) S\left( \frac{1-(d-1)Q}{d} \left(1-|\langle v_0 | p_\mu \rangle| ^{2} \right) \right)
- S \left(1-(d-1)Q \right)
\end{gathered} 
\end{equation}

To maximize $\chi_{BE}$ it is enough to minimize $\langle v_0 | p_\mu \rangle$ under the constraints of eq.(\ref{eq20}) and eq.(\ref{eq21}). The optimization can be done analytically using a parametric representation of $|v_0\rangle,|p_u\rangle,|v_u\rangle$ in 3-D space \cite{branciard2008upper}. This yields:

\begin{equation} \label{eq33} 
\begin{gathered}
\langle v_0 | p_\mu \rangle = e^{-\mu/2}\sqrt{V}-\sqrt{ 1-e^{-\mu} } \sqrt{1-V}
\end{gathered} 
\end{equation}

Equations \ref{eq32} and \ref{eq33} provide the maximal entropy Eve can extract from the system as a function of the QBER $Q$, the visibility $V$, the occupation $\mu$ and dimension size $d$.

\subsubsection{Reselience to noise} \label{Reselience to noise}

\begin{figure}[ht!]
\begin{centering}
\includegraphics[width=0.5\columnwidth]{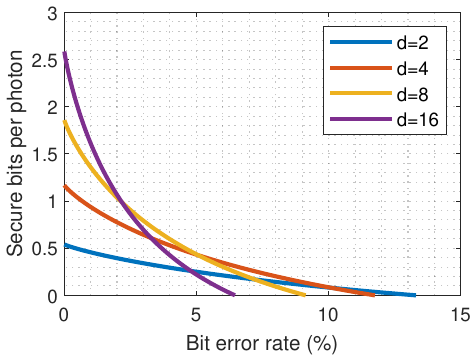}
\par\end{centering}
     \caption{ \textbf{Secure key rate per photon as a function of the bit error rate for dimensions d = 2, 4, 8, 16.} The two-dimensional case is the most robust to noise. Increasing the dimension decreases the maximal QBER which allows a positive secure key rate per photon.}
 \label{Res_to_noise}
 \end{figure}

To test the resilience of our protocol to noise, we calculate the secure key rate per photon versus the bit error rate for different dimensional encoding sizes as presented in Fig. \ref{Res_to_noise}. For $d=2$ we were able to extract a secure key rate up to QBER of $15.4\%$, while for $d=16$ the maximal QBER the protocol could tolerate was reduced to $2.7\%$. This is caused by the linear scaling of the error rate with the dimension due to the leakage of the modulator and the dark counts. Our protocol is, therefore not optimal for increasing communication distances. Fortunately, however, our protocol is useful in many commercially relevant scenarios since in realistic systems the typical error rate is lower than a few percent \cite{xu2020secure}.

\subsection{Detailed photon detection rate analysis}
\label{sec:rate_per_second_analysis}

One of the caveats of our $d$-dimensional qudits is that it extends each qudit over $d$ time bins. In practice, the expectation time per received qudit is limited by the deadtime of the detector. To calculate the received bit rate as a function of the deadtime of the detectors, we write the expectation value for the number of detection events in time window $\Delta t$ as $clicks(\Delta t) = r \Delta t$, where $r$ is the detection rate. Assuming detection events are uncorrelated, 
we can express $clicks(\Delta t)$ by $clicks(t) = P(click)(1+clicks(\Delta t-T-\tau)) + (1-P(click))clicks(\Delta t-\tau)$, where $T$ is the detector deadtime, $\tau$ is the pulse duration and $P(click)$ is the probability a qudit will be recorded by the detector. For detector efficiency $\xi$, dimension $d$ and occupation $\mu$, $P(click)= \frac{\xi \mu}{d}$. We, therefore, get that 
\begin{equation} \label{eq34}
r \Delta t = \frac{\xi \mu}{d}(1+r(\Delta t-T-\tau)) + (1-\frac{\xi \mu}{d})r(\Delta t-\tau)
\end{equation}
from which we can extract the received detection rate $r$:

\begin{equation} \label{eq35}
r = 
\frac{\frac{\xi \mu}{d}}{(T+\tau)\frac{\xi \mu}{d} + \tau(1-\frac{\xi \mu}{d})} = \frac{1}{T+\tau\frac{d}{\xi \mu}}
\end{equation}

The lower bound for the secure key rate generation per second is given by the secure rate per block  (eq. \ref{coherent_1}), which is the outcome of the optimization problem, divided by the average block measurement time 
\begin{equation} \label{photons_per_sec}
T_{meas}=Tp_{det}+\tau(d+1)    
\end{equation}
where $p_{det}$ is the probability that Bob detects a single photon in the block, and $d+1$ time bins are required to encode a qudit and separate the blocks from each other. 

\subsection{Coherent attacks} \label{sup_coh}

\subsubsection{Measurement Description} \label{Sup_Measurement_Description}

The information known by Alice and Bob about the state $\rho_{AB}$ can be represented by known expectation values of two types of projection operators, which correspond to data line measurements and coherence measurements.

For a data line measurement, where a single photon is detected in temporal mode $t=1,2,...,d$, the corresponding measurement operator $M_t$ is defined as:

\begin{equation} \label{SC1}
M_t=\epsilon(1-\epsilon)^{d-1}|\mathrm{vac}\rangle\langle \mathrm{vac}|+(1-\epsilon)^{d}|t\rangle\langle t|
\end{equation}

Here, $\epsilon$ represents Bob's detector's dark count probability, and $|t\rangle$ denotes a single photon state at time-bin $t$.

In addition to data line measurements, Bob performs coherence measurements on successive pulses using the monitoring line. We use the outcome label $k=(c,\pm)$ to represent the coherence measurements, where $c=2,...,d$ denotes the time bin of the last of the two interfering pulses, and $\pm$ represents the bright or dark detector. The corresponding measurement operators are given by:

\begin{equation} \label{SC1}
M_{c,\pm}=\epsilon(1-\epsilon)^{d-1}\ket{\mathrm{vac}}\bra{ \mathrm{vac}}+(1-\epsilon)^{d}\ket{\chi^{\pm}_c}\bra{\chi^{\pm}_c}
\end{equation}

Here, $\ket{\chi^{\pm}_c}= \frac{1}{\sqrt{2}}(\ket{c} \pm \ket{c-1})$ represents the superposition state formed by mode $c$ and mode $c-1$.

\subsubsection{The shared state} \label{The shared state}
The signal states and performed measurements in our protocol can be described by operators on an infinite-dimensional Fock space for each time pulse in Bob's state. However, in order to obtain a numerical lower bound using Eq. \ref{coherent_3}, it is necessary to reformulate the problem in a finite-dimensional manner. For a single block of size $d$, Alice's state can be encoded in a $2^d$ vector space, where a two-dimensional representation indicates whether an occupied pulse was sent or not at each time bin $t\in {1,\ldots,d}$.

Bob's state consists of $2+d$ states: $|\mathrm{vac}\rangle$ represents no photons in the block, $|t\rangle$ corresponds to a single photon at time $t$ and no photons elsewhere, and $|aux\rangle$ denotes two or more photons in the block.

In this protocol, Alice sends an occupied pulse at each time bin with a probability of $1/d$, resulting in an average of one occupied pulse per block. Therefore, the shared state, without an eavesdropper and before considering losses and errors in the channel, is given by $\rho_{AB} = \ket{\psi}\bra{\psi}$, where
\begin{equation}
\ket{\psi} = \sum_{i_1,\ldots,i_d\in {0,1}}\sqrt{\left(\frac{1}{d}\right)^N \left(\frac{d-1}{d}\right)^{d-N}}\ket{i_1,\ldots,i_d}_A \sum_{\tilde{t}} \braket{\tilde{t}}{\sqrt{\mu}^{i_1},\ldots,\sqrt{\mu}^{i_d}}\ket{\tilde{t}}_{B}
\end{equation}
Here, $N=i_1+\ldots+i_d$ represents the number of occupied pulses in the block, and $\tilde{t} \in \{\mathrm{vac},1,\ldots,d, \mathrm{aux}\}$ denotes Bob's states corresponding to no detected photons, a single detected photon, and two or more detected photons in a block.

\subsubsection{Reduced density matrix}
Since Eve's interactions are restricted to Bob's system, the reduced density matrix $\rho_A = \mathrm{tr}_{BE}(\rho_{ABE})$ remains fixed and is determined by the source state. To incorporate this information, we can include the expectation values of $T_k\otimes I$, where $T_k$ represents a complete tomographic operator set on Alice's system.

In particular, since each pulse can be either occupied or vacuum, Alice's state can be represented as a collection of $d$ two-dimensional systems. Any $d$-qubit state $\rho_A$ can be expressed as:

\begin{equation} \label{SC1}
\rho_A = \frac{1}{2^d} \sum_{i_1,\ldots,i_d=0}^{3} S_{i_1,i_2,\ldots,i_d}\hat{\sigma}{i_1}\otimes\hat{\sigma}{i_2}\otimes\ldots\otimes\hat{\sigma}_{i_d}
\end{equation}

Here, $\hat{\sigma}_i$ represents the Pauli matrices, and $S_i$ are the Stokes parameters for multiple qubits. These equations provide us with $4^d-1$ constraints that completely determine Alice's state.

\subsubsection{Sifting maps} \label{SUP sifting}

The sifting process is obtained by announcements that are presented using maps $\Lambda$ that operate on the quantum state. Bob's announcement of registering a single photon event is presented as a map $\Lambda^B(\rho) = F^B\rho F^{B^{\dag}}$ that maps the state to the logical space $\Bar{B}$. $F^B$ is then given by:
\begin{equation} \label{SC3}
F^B = \sum_{t=1}^{d}\ket{t}_{\Bar{B}B} \bra{t}.
\end{equation}
Similarly, Alice's announcement can be modeled as a map $\Lambda^A(\rho) = F^A_v\rho F^{A^{\dag}}$ with:
\begin{equation}
F^A= \sum_{t=1}^{d}\ket{t}_{\Bar{A}A} \bra{0,0,\ldots,1_t,0,\ldots,0}.
\end{equation}

To determine the phase error $\Bar{\delta}$, both parties project the output signals onto the discrete $d$-dimensional Fourier basis states $\ket{t_x}_{\bar{A}}$ and $\ket{t_x}_{\bar{B}}$. In the discrete Fourier basis, the outcomes should be anti-correlated, i.e., $\Bar{t}_x=\mod(d-t_x,d)$. The average phase error $\bar{\delta}$ is given by:

\begin{equation} \label{SC3}
    \bar{\delta} = \tr[F_{\bar{\delta}} \rho_{AB}] =\sum_{v\in V} p(v) \delta_v = 
    G-\tr[F_{x} \rho_{AB}],
\end{equation}
where 
\begin{equation}
    F_{x} = \sum_{t=1}^d (F^{A \dagger} \ket{\bar{t}_x}\bra{\bar{t}_x}F^A)\otimes(F^{B \dagger} \ket{{t}_x}\bra{{t}_x}F^B)
\end{equation}

\subsubsection{Channel model} \label{Channel model}

We describe the losses in the channel using a beam splitter with a transmittance $\eta_{\text{channel}}$, which is connected to an inaccessible environment space. In addition to the channel transmittance, the overall system transmittance is determined by the product of the channel transmittance and the detector efficiency, given by $\eta_{\text{sys}} = \eta_{\text{channel}} \eta_{\text{det}}$.

The probability that Bob detects a photon in the wrong time bin is characterized by an error probability $e_t$, and assumed to be constant and independent of the dimension. This effect is modeled by a completely positive trace-preserving map that incoherently flips each pair of successive pulses with a probability of $e_t$.

Last, we assume an error probability $e_m$ that the outcome interference of two coherent pulses in the beamsplitter exits through the wrong output port. This error leads to the reduction of visibility in the monitoring line.

\subsubsection{Numerical simulation}

In our implementation of the semi-definite program we employ the
packages YALMIP \cite{lofberg2004proceedings} and MOSEK \cite{andersen2000mosek}. 

\subsubsection{Lower bound of the secure key per block} \label{Lower bound of the secure key per block}

We investigate the secure bits per block obtained using equation \ref{coherent_1}. To optimize the lower bound, we vary the signal occupation $\sqrt{\mu}$ of Alice's source, typically on the order of 0.1. As expected, we observe that higher-dimensional encoding sizes lead to a reduction in the secure key per block. Moreover, the secure bits per block exhibit a quadratic decrease with the increasing loss for encoding sizes in any dimension. The secure bits per second are presented in the main text.

 \begin{figure}[ht!]
\begin{centering}
\includegraphics[width=0.5\columnwidth]{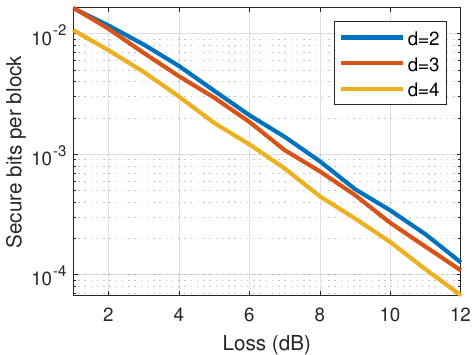}
\par\end{centering}
     \caption{ \textbf{Lower bound of the secure key per block}. The Lower bound on the secure key rate per block versus the system loss in dB for the high dimensional COW with nearest neighbors interference protocol is calculated using Eq. \ref{coherent_1}. Different colors represent different dimensional encoding sizes. All curves exhibit a quadratic decay with the loss, where the optimum is achieved for $d=2$.}
 \label{Fig secure bits per block}
 \end{figure}

\subsection{Experimental details} \label{Sup_Experimental_details}

\subsubsection{Key rates for different dimensions} \label{Key rates for different dimensions}

\begin{figure}[ht!]
\begin{centering}
\includegraphics[width=0.5\columnwidth]{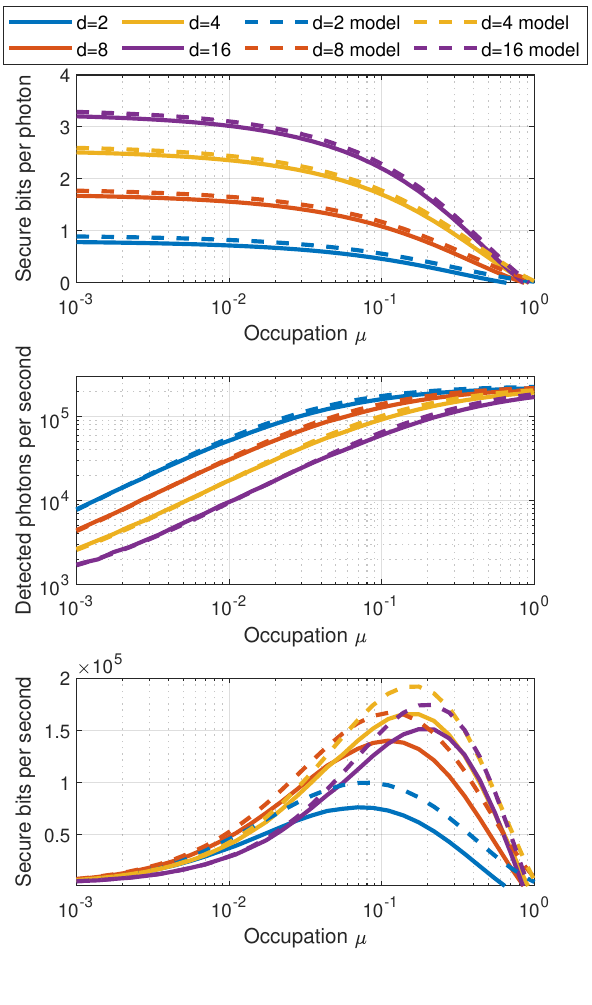}
\par\end{centering}
     \caption{ \textbf{Key rates for different dimensions.} a) Number of secure bits per photon (solid) for our system and for the model system (dashed). Increasing the pulse occupation weakens the constraints on Eve and therefore increases the information she can obtain, yielding a lower number of secure bits per photon. Higher dimensional encoding allows a higher number of secure bits per photon. b) Number of raw bits per photon in our system (solid) and the calculated raw bits per second for the model system. c) Secure bits per second is multiplying the raw bits per photon by photons per second. The optimum is achieved at $d=8$, where the number of secure bits per second increases by $2.04$ for our system, and by $1.89$ for the model.}
 \label{Fig3}
 \end{figure}

We first calculate the number of secure bits per photon as a function of the pulse occupation for different dimensions, as presented in Fig. \ref{Fig3}a. Solid lines present the number of secure bits per photon for our system, based on the measured QBER and visibility for each dimensional encoding size and Eq. \ref{upper_bound}. The dashed lines present the calculated secure bits per photon for the theoretical model, where we assumed visibility of $99\%$ and a QBER of $0.4\%$ per time bin. Here we also assumed the QBER scales linearly with the dimension size. The main source for such linear scaling is the finite extinction ratio of the intensity modulator, typically on the order of  $0.01$. In the limit of low occupation, the detector's dark counts may become the dominant source for linear scaling of the QBER with the dimension size.

As appears in Fig. \ref{Fig3}a, higher dimensional encoding allows higher secure bits per photon. At the same time, increasing the pulse occupation weakens the constraints on Eve and therefore increases the information she can obtain, decreasing the number of secure bits per photon. In Fig. \ref{Fig3}b we present the number of detected photons per second versus the pulse occupation, for different dimensions. The solid lines present the measured number of detected photons per second, and the dashed lines present the calculated detection rate (see supplementary information \ref{sec:rate_per_second_analysis}): $\frac{1}{T+\tau\frac{D}{\xi \mu}}$. The number of raw bits per photon increase linearly up to an occupation of around $0.05$ where the detector starts to saturate.

\subsubsection{Errors sources} \label{Error sources}

\begin{figure}[ht!]
\begin{centering}
\includegraphics[width=0.5\columnwidth]{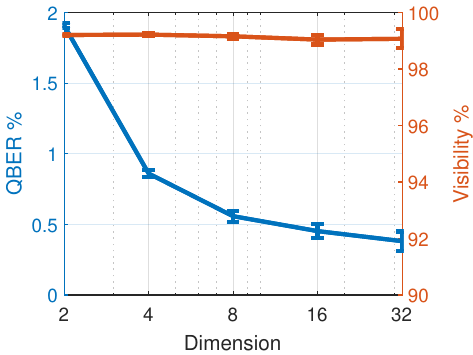}
\par\end{centering}
     \caption{ \textbf{Bit error rate and visibility as a function of the protocol's dimension.} The QBER per encoding time bin decreases with the dimension size, due to electronic ringing common in high-rate modulation systems. The measured visibility is insensitive to the dimension size. Error bars are calculated assuming shot-noise limited detection.}
 \label{fig4}
 \end{figure}

In practice, one of the main noise sources in fast modulation transmitters is cross-talk between successive pulses, due to electronic ringing. Since higher dimensions result in longer average times between successive pulses, the sensitivity to cross-talk between successive pulses decreases with the dimension size. In our system, we experimentally observed that the QBER per encoding time bin decreases as the dimension is increased (Fig. \ref{fig4}), yielding higher secure bit rates than the model's prediction. Another source of noise is the detector's dark counts which is in our case less than 5000 and therefore does not play an important role in the saturated regime quantum key distribution.

\end{document}